\documentclass[journal]{IEEEtran}
\usepackage{bbm}
\usepackage{graphicx}
\usepackage{epsfig,subfigure}
\usepackage{amssymb,amsmath}
\usepackage{enumerate}
\usepackage{latexsym}
\usepackage{booktabs}
\usepackage{multirow} 
\usepackage{array} 
\usepackage{cite}
\usepackage{comment}
\usepackage{todonotes}
\usepackage{authblk}
\usepackage{url}
\usepackage{algorithm}
\usepackage{algorithmic}
\usepackage{caption}
\usepackage{hyperref}
\newtheorem{thm}{Theorem}

\renewcommand{\vec}[1]{\boldsymbol{#1}}
\newcommand{\vx}{\vec{x}}
\newcommand{\vy}{\vec{y}}

\newcommand{\cX}{\mathcal{X}}

\usepackage{graphicx}
\graphicspath{{Figures/}}

\begin{document}
\title{RSU-Based Online Intrusion Detection and Mitigation for VANET}
\author{Ammar Haydari, 
Yasin Yilmaz
\thanks{
Authors are with the Electrical Engineering Department, University of South Florida, Tampa, FL USA (e-mail: ammarhaydari@mail.usf.edu, yasiny@usf.edu). }
}

\maketitle

\begin{abstract}

Secure vehicular communication is a critical factor for secure traffic management. Effective security in intelligent transportation systems (ITS) requires effective and timely intrusion detection systems (IDS). In this paper, we consider false data injection attacks and distributed denial-of-service (DDoS) attacks, especially the stealthy DDoS attacks, targeting the integrity and availability, respectively, in vehicular ad-hoc networks (VANET). Novel statistical intrusion detection and mitigation techniques based on centralized communications through roadside units (RSU) are proposed for the considered attacks. The performance of the proposed methods are evaluated using a traffic simulator and a real traffic dataset. Comparisons with the state-of-the-art solutions clearly demonstrate the superior performance of the proposed methods in terms of quick and accurate detection and localization of cyberattacks. 
\end{abstract}

\section{Introduction}
\label{s:intro}

Improving transportation safety is one of the main research areas for intelligent transportation systems (ITS) \cite{hasrouny2017vanet}. An important facilitator for secure and reliable traffic flow is the data dissemination through Vehicular Ad-Hoc Network (VANET), including vehicle to vehicle (V2V) communications and vehicle to infrastructure (V2I) communications. VANET is a promising technology that enables communications between driverless autonomous vehicles, which are expected to dominate the future traffic, as well as traditional vehicles controlled by a driver \cite{amoozadeh2015security}. VANET applications can be classified into two types as traffic safety applications and traffic management applications. Route planning applications for drivers is an example for the traffic management applications. The safety-related applications are exemplified by road condition applications and accident information systems. 

There are two possible communication methods for VANET: (i) 5.9 GHz dedicated short range communication (DSRC), and (ii) cellular-based vehicular communication \cite{uhlemann2018battle}. With fast advancing 5G technologies for connected and automated vehicles, industry is more inclined to support on cellular based communication technologies, named as cellular vehicle to everything (C-V2X) \cite{mannoni2019comparison}. Recently, Europe and the US both announced to advance their technologies for cellular-based VANET. Although cellular network models are more decentralized, still such C-V2X communication requires RSU or some sort of base stations near the roads in order to collect and process traffic and vehicular data \cite{lee2019adaptive}. 


In VANET, different types of data such as position information, road conditions and emergency messages are disseminated. The availability and integrity of such data are the two essential aspects of VANET security. DSRC-based ad hoc communication in ITS requires distributed roadside units (RSUs), which have a critical role in VANET as a static infrastructure over the roads for centralized communication. RSUs provide high connectivity and security in traffic, which is feasible especially in high-demand urban areas which are highly vulnerable to cyberattacks. 
Considering the potential life-threatening outcomes in traffic, cyberattacks to VANET need to be quickly and accurately detected and mitigated. In DSRC, while vehicles collect broadcasted messages to take decisions, RSUs collect those messages for traffic management purposes, 
and securing the VANET communications for V2V and V2I. To this end, in this paper, we propose a centralized RSU-based statistical detection and mitigation method for cyberattacks targeting both data availability and integrity. 

\subsection {False Data Injection Attacks Targeting Integrity}

Falsified message content may cause the drivers to take wrong actions entailing devastating and life-threatening results to the vehicular traffic. Autonomous vehicles are exposed to an even greater risk due to false data injection (FDI) attacks as their automatic decisions may rely more on the received VANET messages. For example, position is one of the most important information in VANET; when a vehicle sends wrong position information, then a nearby autonomous vehicle may accelerate according to the received falsified message. An effective intrusion detection system (IDS) should effectively deal with FDI attacks, in which attacker sends bogus information to the network in order to change the vehicle behaviors in traffic. Once an intruder injects bogus data to the network, it should be detected and mitigated timely to prevent a major problem such as an accident or traffic congestion. 

There are several detection approaches for different FDI attack models. Trust-based security mechanisms and behavior-based security mechanisms are two common signature-based detection approaches for FDI attacks in the literature. However, they are mostly not computationally efficient, and cannot detect new attack patterns that do not conform to the known signatures \cite{soleymani2015trust}, \cite{khan2015detailed}. In this work, we propose a statistical anomaly detection method that can quickly detects FDI attacks, including the previously unseen ones, as opposed to the signature-based methods. Our method is implemented on RSU, and it monitors the data stream received from each vehicle within its communication range. We do not use any revocation list or voting list scheme. Once our method detects an anomalous vehicle, it blocks the data transfer from that vehicle and informs the other vehicles.

\subsection {DDoS Attacks Targeting Availability}

Availability of the communications is one of the main objectives in ITS. Denial-of-service (DoS) attacks target the availability of network service, e.g., by sending high volume (flooding) of data packets to the service provider. Once a DoS attack is launched successfully on VANET, e.g., on RSU, the system operation shuts down such that no one can get regular service. Unavailability of the VANET service due to a DoS attack may cause a significant damage to the vehicular traffic. Compared to the FDI attack, it is easier to initiate a DoS attack for attackers as no data manipulation is needed; however, the FDI attack poses a bigger threat since wrong data is usually more detrimental than no data. In practice, to make the mitigation more difficult, attackers synchronously launch a DoS attack from multiple sites, which is called a Distributed DoS (DDoS) attack. The proliferation of Internet-of-Things (IoT) devices, in particular autonomous vehicles, facilitates a new type of stealthy DDoS attack, called Mongolian DDoS attack \cite{nexusguard}, or low-rate DDoS \cite{zhang2012flow}, \cite{chen2018power}, that can easily bypass traditional IDSs such as data traffic filters and firewalls while still causing a significant disruption in the targeted service due to its highly distributed and synchronous nature. 


It is quite challenging to timely detect and mitigate stealthy DDoS attacks compared to the standard brute-force DDoS attacks because the increase in the individual data rates from multiple parties with respect to their nominal baselines can be very low such that traditional data filtering methods cannot detect them. Yet, the aggregate increase in the data traffic received by the targeted RSU from multiple attackers can be tremendous, thus the RSU gets overwhelmed and stops serving legitimate users. 
There are three main enabling factors that make such stealthy DDoS attacks relevant in VANET:

\emph{Increasing number of connected vehicles:} In near future, connected vehicles will be widespread, which will in turn increase the attack surface for DDoS attacks. Similar to Internet of Things (IoT), or as a part of it, this trend is called Internet of Vehicles (IoV). Growing attack surface will enable effective DDoS attacks with less and less increase in data rate (i.e., less and less attack signature) and more and more stealthy nature over time. 

\emph{Public V2I communications:} Since V2I communications is open to everyone, attackers can utilize many other IoT devices that pretend to be vehicles to spoof the RSU. Such non-vehicle attacking devices can be on vehicles or static within the communication range of RSUs. The possibility to use IoT devices for attacking RSU greatly extends the attack surface, making stealthy DDoS a big threat for VANET. 

\emph{Sybil attack:} Furthermore, through Sybil attack the attackers do not even need to utilize many devices (vehicle or non-vehicle) for an effective stealthy attack. In Sybil attack, a single vehicle pretends to be multiple vehicles by creating fake identities. Hence, a single vehicle can transmit a huge amount of data packets to the RSU appearing as coming from different vehicles at low transmission rates. This efficient attack strategy can generate stealthy DDoS attacks by using only a fraction of the number of vehicles/devices that is normally needed.

In this paper, we propose a powerful multivariate-statistical method for the timely detection and mitigation of stealthy DDoS attacks to RSU. 

\subsection {Contributions}

In this paper, we propose a novel statistical anomaly-based detection and mitigation technique to address FDI attacks and flooding-based (stealthy or brute-force) DDoS attacks targeting VANET, in particular RSU. Our contributions can be summarized as follows.
\begin{itemize}
\item A novel statistical detector is proposed for FDI and DDoS attacks implemented on RSU.
\item The asymptotic false alarm rate of the proposed detector is analyzed, and a closed form expression for the detection threshold is derived based on this analysis.
\item Based on the detection method, a statistical anomaly localization, and accordingly attack mitigation method is proposed for FDI and DDoS attacks for VANET.
\item Performance of the proposed detection and mitigation methods are extensively evaluated using state-of-the-art traffic simulators and a real traffic dataset. To the best of our knowledge, this is the first work to use real traffic data in the cybersecurity literature for VANET. 
\end{itemize}

The rest of the paper is organized as follows. Related works are discussed in Section 2. The traffic and attack models for the considered attack types are given in Section 3. The proposed statistical detection and mitigation methods are presented in Section 4. Numerical results are provided in Section 5. Finally, the paper is concluded in Section 6. Throughout the paper, lowercase and uppercase bold letters are used to denote vectors and matrices, respectively. 

\section{Related Work}
 
\subsection{False Data Injection Attacks}
 
Injection of fake messages is a high threat to ITS/VANET security \cite{6899663}. There are several key features that differentiate VANET and ITS security from other network security topics, such as high mobility, dynamic characteristics, and life-threatening conditions. One popular type of FDI attack is misbehavior detection where data is compared with the behavior of vehicles. Many misbehavior detection models are proposed in the literature considering either the trust-centric or data-centric approach for enhancing the security of VANETs \cite{wahab2014cooperative, li2015reputation, van2018survey}. However, these models mostly 
rely on the data content, which makes them not generalizable. Trust-centric models are based on voting or scoring schemes, in which the reliability of a node broadcasting a message is voted by the other nodes receiving the message. Once the cumulative voting score exceeds a level against the node, it is declared as intruder, and its message dissemination is blocked \cite{wex2008trust, hasrouny2019misbehavior}. General data-centric approach evaluates the driver's behavior with respect to shared messages. In \cite{ruj2011data}, an example data-centric detection model is proposed for emergency messages, where behavior of the node and the received message are compared. The authors proposed a malicious vehicle detection mechanism using specific alert messages that convey information about the emergency of traffic or emergency of the vehicle. If there is a mismatch between the received packets and the behavior of the node transmitting the packets, system raises a false data alarm, and informs the other nodes about the false data and misbehaving node. In \cite{sedjelmaci2014efficient} and  \cite{sedjelmaci2017predict}, intrusion detection mechanisms were proposed against multiple attack types including FDI and packet drop attacks using vehicle reputation scores collected by RSUs. 

There are several misbehavior detection mechanisms for VANET based on statistical and machine learning approaches. Authors in \cite{ghaleb2019hybrid} proposed a decentralized misbehavior detection mechanism using context-reference model with Kalman and Hample filters to extract the consistency of message with the behavior of vehicles. Another statistical misbehavior detection model is proposed in \cite{sharshembiev2019fail} using entropy-based classification. 
In \cite{zaidi2016host}, authors proposed a statistical anomaly detection technique considering only the data content instead of trust score or revocation list to detect anomalous nodes that inject falsified data into VANET. This paper uses the Greenshield traffic model assuming close vehicles have similar flow, speed, and density values. In a decentralized traffic, each vehicle calculates its own value and compares it with average received values until the average value is below the predefined threshold. After that, a t-test is applied to these values to determine if the received message comes from an intruder or not. Recently a new security mechanism for VANET is proposed in \cite{liang2019filterVanet} with a novel detection and classification mechanism by using a hidden generalized mixture transition distribution model. In this work, detector collects the data of each vehicle in feature tables, and classifier processes the extracted knowledge. Since the proposed detection models run on network level by examining the VANET as a whole, collaborative FDI attacks decreases the performance of the proposed detection algorithm. 

Anomaly detection mechanisms based on machine learning provide suitable results for security mechanisms in general. 
Recently, several machine learning based IDS models are studied for ITS/VANET \cite{eziama2018malicious, gyawali2019misbehavior, so2018integrating, so2019physical, singh2019machine}. In \cite{eziama2018malicious}, a trust-based cyberattack detection model is proposed for false position, timing, and Sybil attacks using Bayesian deep learning. A security model based on plausibility check is studied using machine learning in \cite{so2018integrating}, where authors considered supervised $k$NN and SVM algorithms for detecting anomalies on feature vector by experimenting with the VaReMi dataset \cite{van2018veremi}. An improved version of this work by introducing new plausibility check methods is presented in \cite{so2019physical}. Although this is not a centralized detection model, it still requires some information from RSU or a trustworthy source for calculating the plausibility feature vector. In \cite{singh2019machine}, supervised learning models are used for detecting false position attacks in VANET. A cooperative misbehavior detection model is proposed in \cite{gyawali2019misbehavior} for detecting emergency message falsification and position falsification attacks.

\subsection{Denial of Service Attacks}
  
DoS attacks may cause catastrophic effects to the vehicular traffic since the decisions of autonomous vehicles may critically depend on the communications between the vehicles and the infrastructure \cite{7872388}. There are several solution methods proposed for DoS attacks in the literature. In \cite{soryal2013attack}, a statistical DoS attack detection model was proposed for the IEEE 802.11 DCF protocol. A Markov chain-based adaptive threshold was used for received Clear-to-Send (CTS) packets. If the received CTS rate is above the threshold, the source node is labelled as attacker. Another DoS attack detector was studied in \cite{verma2013prevention}, where MAC layer ACK/SYN packets rates are monitored and compared with a predefined threshold at a centralized node.
These methods are not suitable for detecting stealthy DDoS attacks as they directly compare the observed packet rate to a threshold. 
  
A packet delivery ratio based jamming attack detection model for VANET was presented in \cite{mokdad2015djavan}, where two different traffic scenarios were considered for performance evaluation. In \cite{kerrache2017tfdd}, authors developed a trust-based framework named TFDD, which is a hybrid detection model based on signature-based and anomaly-based detection models. It uses a data verification module with honesty weight and quality weight in order to detect attacking vehicles. An unsupervised detection method based on the k-means clustering algorithm was proposed in \cite{karagiannis2018jamming} for jamming attacks in RF-based vehicular communication. A recent work \cite{lyamin2018ai} on jamming attacks for platooning vehicles in VANET studies a hybrid detection model with a statistical approach. This detection model focuses only on platooning vehicles to detect jamming attacks, and is not applicable to other DDoS attacks. A recent DDoS detection model for VANET is proposed in \cite{kolandaisamy2018multivariant}, where authors use a multivariant stream analysis (MVSA) approach. Since this IDS follows a window-based approach, the size of time windows limits its timely detection performance. In \cite{haydari2018real}, we proposed an anomaly-based IDS for mitigating DDoS attacks in VANET, which is significantly enhanced with theoretical and experimental results, and extended to FDI attacks in this paper. 
 
There are some other security mechanisms proposed for other attack types. Black hole attack or packet drop attack in which packets are deliberately dropped at a compromised node is another type of DoS attack that is studied in the VANET literature. For instance, \cite{wahab2016ceap} proposed an SVM-based detection method for clustered VANET. In Sybil attack, attacker identifies itself as multiple nodes. For example, in \cite{yu2013detecting}, authors considered detecting Sybil attacks using strength analysis for received signals with statistical position verification.
 
\subsection{Comparisons}
There are several novelties in our work compared to the existing methods in the literature. 
Firstly, the existing IDS models consider several attack strategies which only target one security objective, e.g., IDS for data integrity. Whereas, our proposed security model addresses two distinct attack types, FDI and DDoS. 

Regarding the FDI attacks, the studies discussed up to now consider network level IDSs, which is not suitable for detecting collaborative FDI attacks with multiple attackers. Whereas, our proposed anomaly-based IDS at RSU classifies each vehicle as malicious or benign, hence is robust to collaborative FDI attacks. Once detected, such an attack is easily mitigated by dropping packets from attackers without interrupting the benign communications. Another strength of our IDS is that vehicles only need to broadcast their own messages; they do not need to have information tables for neighboring vehicles. Other works in the literature also typically assume a specific message type or data content. Since, our proposed model is a statistical IDS, it is not restricted to any message format and data content. After it detects the FDI attack, it also identifies the specific data type which is under attack, such as speed, position or emergency message. 

Regrading the DDoS attack, our proposed method is advantageous compared to the existing works in several ways: (i) it does not consider any specific data model for DDoS attacks, (ii) it successfully detects highly distributed DoS attacks from many attackers, and (iii) it can handle both stealthy and traditional brute-force DDoS attacks. Other studies in the literature for DoS attacks in VANET do not consider distributed or stealthy DoS attack scenarios in their experiments.
Our method can timely detect stealthy DDoS attacks, which is a quite challenging task in the highly dynamic VANET environment. In stealthy DDoS, as demonstrated recently on the Internet \cite{nexusguard}, while slightly increasing the data rate from one source do not effect the regular performance of the receiver node, attacking from multiple sources, even with slightly increasing the transmitted data rate, can cumulatively overwhelm the receiver node. 

Recent machine learning based methods are based on sample-by-sample outlier detection, and do not continuously update their statistics for detecting anomalies, as opposed to our proposed sequential detection method. The sample-by-sample outlier detection approach is known to be prone to frequent false alarms due to nominal outliers \cite{baker2016statisticians} while sequential methods can avoid nominal outliers and detect only the persistent outliers as anomalous \cite{basseville1993detection}.
 \section{System Model}

\subsection{VANET Model}
\label{s:vm}

\begin{figure}[t]
\centering
\includegraphics[width=.5\textwidth]{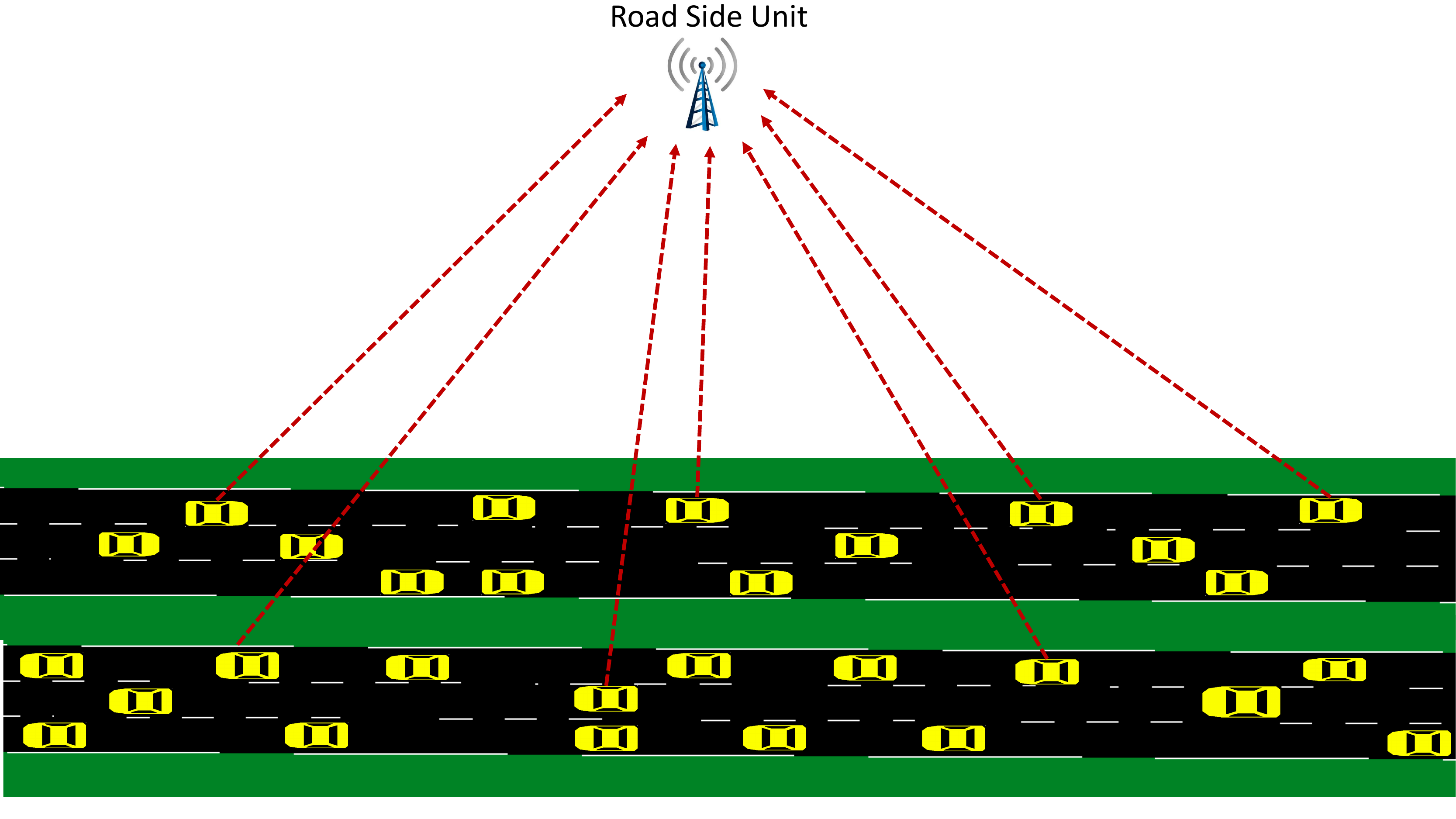}
\caption{Traffic model for the nominal case where all vehicles broadcast messages and RSU collects these messages.}
\label{f:normal}
\end{figure}

A general two-way traffic flow is considered as shown in Fig. \ref{f:normal}. However, the proposed IDSs are not restricted to a specific road type; they can perform well on different scenarios such as one-way traffic, two-way traffic, urban area, highway area etc. Vehicle-to-vehicle (V2V) and vehicle-to-RSU (V2I) communications based on broadcasting take place in the considered VANET model to disseminate beacon messages. In general, such messages may have various content. In this work, we consider that each vehicle regularly broadcasts messages in the (ID, Speed, Position, Direction) format. 

All messages are protected with cryptographic algorithms, but such details are out of the scope of this paper. We assume that different pseudonyms are assigned by a central authority to each vehicle for providing authenticity and identification. Thus, ID of each vehicle is always known by the RSUs. Collected messages can be used for different purposes, but they are mostly used for informing other vehicles on the road. For instance, an RSU calculates the average speed and the density of road using the received beacon messages from the vehicles in its range, and conveys these calculated messages to the other RSUs to inform the vehicles that are not in the range. RSUs play a central role in the security of VANET. Hence, we propose a statistical IDS that runs at each RSU. Although the beacon messages are already encrypted for secure communications, the integrity, i.e., correctness, of message content, as well as the availability of VANET communications should be maintained.

\subsection{Attack Model}

In this work, we consider two types of attacks in VANET. 

\subsubsection{False Data Injection Attacks}

Creating completely a false message or changing some parameters of a message may have a crucial impact on the traffic. Although today's traffic is not dominated yet by fully autonomous vehicles, in the near future it is expected that the majority of vehicles will decide by themselves without human interaction. In such a scenario, disseminating correct messages to other vehicles is a top priority for VANET. For example, a malicious vehicle conveying false messages about its position and speed may cause other vehicles to take wrong actions such as decreasing speed or changing lanes. Even without any malicious intent, faulty sensors in a vehicle may result in false messages. The proposed RSU-based statistical IDS can quickly detect FDI attacks, and accurately identify the false data type and its source vehicle. 

\subsubsection{Distributed Denial-of-Service Attacks}

In the considered DDoS attack, the number of messages per unit time, i.e., data rate such as packets/sec., from multiple sources (vehicles or other devices pretending to be vehicles) increases synchronously \footnote{No strict synchronization is needed to perform a DDoS attack}. Unless quickly detected and mitigated, such a flooding of messages may easily overwhelm the attacked RSUs to make the VANET communications unavailable. It is significantly more challenging to detect and mitigate stealthy DDoS attacks than the traditional brute-force DDoS attacks. The aggregate data rate received by the RSU is still high enough to take it down; however, the increase in individual data rates of attacking nodes is low such that they easily remain undetected by traditional IDS, such as data filters and firewalls. Despite its relatively small increase in the individual data rates, the greatness in the number of attacking nodes is what makes a stealthy DDoS attack threatening. 

The proposed statistical IDS monitors the data rate, thus it is not restricted to the data format defined above. In the following sections, we show that the proposed IDS can effectively handle both stealthy and brute-force DDoS attacks, as opposed to the traditional data filtering methods.

\section{Proposed Intrusion Detection and Mitigation Systems}

In general, anomaly-based IDS works by comparing the observed data instances with the statistical model of nominal operation learned from training data, and possibly also with the anomalous statistical model learned from training data as well. Anomaly-based IDS can be categorized considering three aspects: availability of training data, parameterization of statistical model, and sequential decision making. 

In terms of availability of training data, anomaly-based IDS can be categorized into groups, semi-supervised and supervised. If there is only nominal training data, IDS aims to detect the significant deviations from the learned nominal statistical model, which is called the semi-supervised setting. Whereas, in the supervised setting, IDS builds also a statistical model for the considered attacks using available data instances from previous attack cases, and compares the goodness-of-fit (e.g., likelihood) of the observed data under the nominal and attacks models.  

In terms of model parameterization, there are also two types, parametric and non-parametric methods. While parametric methods try to fit certain parametric probability distributions (e.g., Gaussian, Poisson, etc.) to the data, non-parametric methods try to learn statistical patterns from data without assuming certain probability distributions (e.g., distance-based and histogram-based methods). 

Finally, in terms of sequential decision making, we also have two groups: sample-by-sample outlier detection methods and sequential anomaly detection methods. Outlier detection methods decide for each observed data instance as either nominal or anomalous. However, sequential methods updates a decision statistic using each observed data instance, and decides for anomaly when there is enough statistical evidence. Mathematically, the objective of sequential methods is to minimize the expected number of data instances used to detect anomaly while satisfying a constraint on false alarm rate. 

\subsection{Challenges in VANET for an IDS}
\label{s:challenges}

Implementing an anomaly-based IDS in VANET is a challenging task due to several reasons. Three major challenges in VANET for an IDS can be summarized as follows.

\begin{itemize}
\item[(C1)] \emph{Unknown attack patterns:} As opposed to the traditional computer networks and the Internet, the possible attack patterns (i.e., signatures) are mostly unknown in the emerging field of ITS/VANET security. Hence, conventional signature-based IDS, which can only detect the known attack signatures, and supervised anomaly-based IDS are in general not suitable for VANET. 

\item[(C2)] \emph{Disparate data types:} Since anomalies occur relative to the context defined by the entire data dimensions, they should ideally be jointly monitored through multivariate analysis. However, due to the disparate data types conveyed in messages, the multivariate probability distribution of the message content is quite complicated. For instance, speed data is numerical, direction is angular, and position is numerical/angular. As a result, parametric anomaly detection methods, which try to fit tractable probability distributions to the training data, are not feasible here.

\item[(C3)] \emph{Timely and minimally invasive mitigation:} Considering the life-threatening and economic concerns of a failure in VANET communications, cyberattacks should be quickly mitigated in a minimally invasive manner. The identification of malicious users should also be accurate such that the legitimate users continue receiving regular service. It is known that sequential methods are much more effective in timely detection than sample-by-sample outlier detection methods \cite{poor2009quickest}.
\end{itemize}

To address the challenges above we propose a statistical IDS based on a semi-supervised, non-parametric and sequential anomaly detection technique. 

\subsection{Proposed Detection and Mitigation Method for FDI attack}
\label{fdi}

In this section, we explain the proposed intrusion detection and mitigation method for FDI attacks. The proposed method is an online intrusion detection system and can work in real-time, similar to a sequential anomaly detection algorithm \cite{yilmaz2017online}. Thanks to its semi-supervised, nonparametric, and sequential methodology, the proposed method addresses well the challenges (C1)-(C3), in Section \ref{s:challenges} respectively.

\begin{figure}[t]
\centering
\includegraphics[width=.49\textwidth]{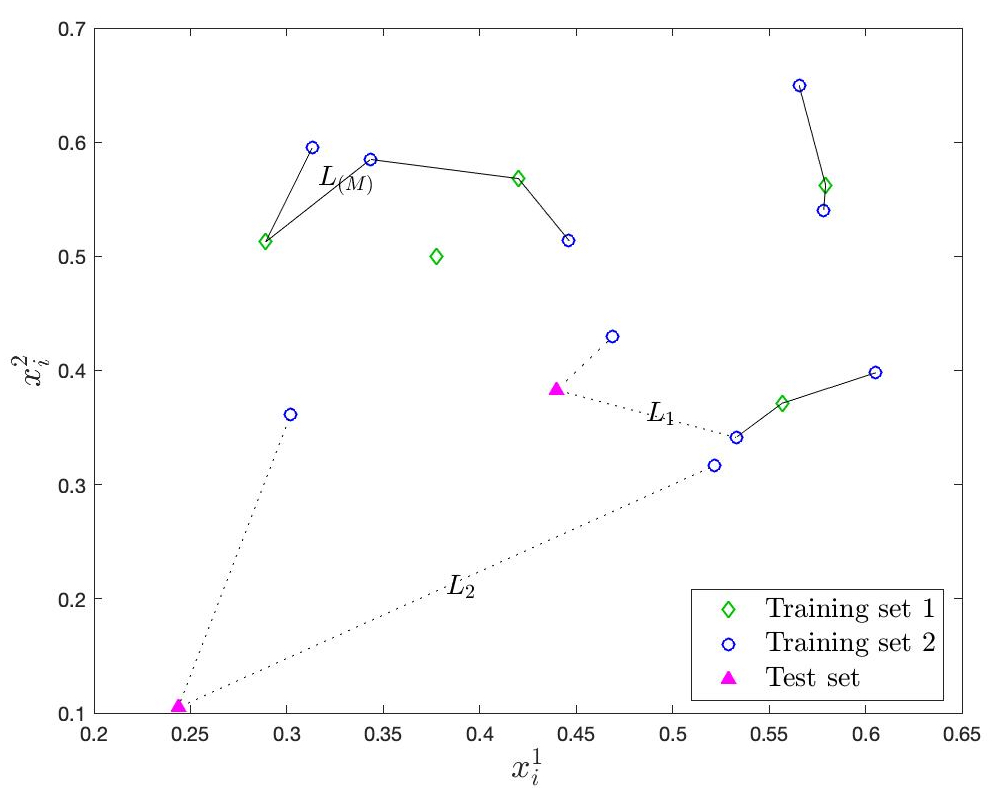}
\caption{Proposed detection procedure with $N_1=5$, $N_2=10$, $M=4$, $k=2$, $s=1$, $\gamma=1$. $(L_1)^d-(L_{(M)})^d$ and $(L_2)^d-(L_{(M)})^d$ are used to update the test statistic $s_t$ and raise an alarm at time $T$ as shown in (\ref{e:length})--(\ref{e:tmin}). Training and test points are generated from a bivariate normal distribution with independent components, $0.5$ mean and $0.1$ standard deviation.}
\label{f:graph}
\end{figure}

In the proposed IDS, each RSU runs a separate detector for each vehicle that it receives messages from. In particular, an RSU starts to monitor a vehicle when the vehicle enters its range until either the vehicle exits the range or an anomaly in the message content is detected.

Consider that each data instance $\vx_i \in \mathbb{R}^d$ is a $d$-dimensional real-valued vector representing the observed $d$ data dimensions $\{x_i^1,\ldots,x_i^d\}$ depending on the application. 
While we consider (ID, Speed, Position, Direction) as the data dimensions, the proposed detection and mitigation methods are not restricted to this particular setup. 
The data instances are inspected by an infrastructure unit (RSU) in a centralized fashion for detecting falsified data attacks. We do not have any assumption on the probability distributions of data dimensions, e.g., they can be correlated or even follow disparate distributions. It is only assumed that each data dimension can be normalized to [0,1] by using the minimum and maximum values, which is needed to deal with the heterogeneity among data dimensions.


In the training phase, firstly the nominal training data $\mathcal{X}_N=\{\vx_1,\ldots,\vx_N\}$ is randomly partitioned into two sets $\mathcal{X}_{N_1}$ and $\mathcal{X}_{N_2}$, where $N_1+N_2=N$, for computational and theoretical purposes. Typically, $N_2$ is selected greater than $N_1$ as explained in Theorem \ref{thm:1}.  
The nominal training data $\cX_N$ is obtained through historical observations in the range of RSU.  Then, for each data point in $\mathcal{X}_{N_1}$,  $k$ nearest neighbors ($k$NN) in $\mathcal{X}_{N_2}$ in terms of Euclidean distance are found. For each point $i$ in $\mathcal{X}_{N_1}$ the total distance $L_i$ is computed as

\begin{equation}
\label{e:length}
L_i = \sum_{j=k-s+1}^k e_{ij}^\gamma,
\end{equation}
where $e_{ij}$ is the Euclidean distance from point $i$ in $\mathcal{X}_{N_1}$ to its $j$th nearest neighbor in $\mathcal{X}_{N_2}$. The weight $\gamma>0$ and the number of considered neighbors $s$, which is a number between $1$ and $k$, are introduced to increase the flexibility of method.

Next, for a significance level $\alpha$, for which a typical choice is $0.05$, the $(1-\alpha)$th percentile $L_{(M)}$ of $\{L_i: i=1,\ldots,N_1\}$ values is found, where $M=\lfloor N_1 (1-\alpha) \rfloor$, and $\lfloor \cdot \rfloor$ is the floor operator. The $L_{(M)}$ value is later used as a baseline in the test to evaluate the anomaly evidence in the test instances. Depending on the VANET structure, if the range of RSU is composed of heterogeneous road segments with different speed and direction baselines, then multiple training sets, and accordingly multiple $L_{(M)}$ values can be obtained for such road segments. During the training phase, actually an Euclidean $k$NN graph is formed between $(1-\alpha)\%$ of the points in $\cX_{N_1}$ with the smallest $L_i$ values and their neighbors in $\cX_{N_2}$, as illustrated in Fig. \ref{f:graph}. As we will show next, in the test phase, we actually evaluate how far/close a test instance is in becoming a vertex in this graph if it were to be included in $\cX_{N_1}$. From another perspective, $(1-\alpha)\%$ of the points with the smallest $L_i$ values in $\cX_{N_1}$ is an estimate of the ``minimum volume set" which is the most compact set that has at least $1-\alpha$ probability \cite{hero2007geometric}, and in the test phase, we measure how far/close a test instance is to be included in this most compact set.

In the test phase, to evaluate the anomaly evidence in a newly observed instance $\vx_t$ at time $t$, we compute how small/big the total distance $L_t$ of $\vx_t$ compared to the baseline $L_{(M)}$, which corresponds to a boundary point in the most compact set of nominal points. Specifically, at each time $t$ we compute the total distance $L_t$ with respect to the nominal training set $\cX_{N_2}$ as in \eqref{e:length}, and the anomaly evidence
 \begin{equation}
\label{e:Dt}
D_t = (L_t)^d - (L_{(M)})^d.
\end{equation} 
Note that the anomaly evidence provided by $D_t$ can be positive or negative. Unlike the sample-by-sample outlier detection methods, the proposed IDS does not decide for each instance, but rather accumulates the anomaly evidences over time by updating its decision statistic as  
\begin{equation}
\label{e:st}
s_t = \max\{s_{t-1} + D_t,0\}, ~s_0=0.
\end{equation}
It decides for an anomaly only when enough evidence is accumulated in the decision statistic, i.e., at time
\begin{equation}
\label{e:tmin}
T = \min\{t: s_t \ge h\},
\end{equation}
similar to the CUSUM algorithm \cite{basseville1993detection}. Note that the proposed IDS in \eqref{e:length}--\eqref{e:tmin} is a novel data-driven algorithm, as opposed to the parametric (model-based) CUSUM algorithm. 

The detection threshold $h$ manifests a trade-off between  minimizing detection delay and minimizing false alarm rate. For example, higher $h$ decreases the false alarm rate at the expense of larger average detection delay, and vice versa for lower $h$. Next, we show how to set $h$ to satisfy a desired false alarm rate. 

\begin{thm}
\label{thm:1}
As the training set grows ($N_2 \rightarrow \infty$), with $k=s=\gamma=1$, to asymptotically ensure that the false alarm rate is less than or equal to a desired level $\beta$, the threshold $h$ can be chosen as follows
\begin{equation}
\label{e:thm}
h = \frac{-\log \beta}{\omega_0}.
\end{equation}
In \eqref{e:thm}, $\omega_0>0$ is given by
\begin{align}
    \omega_0 &= v_d - \theta -\frac{1}{\phi} \mathcal{W}\left( -\phi \theta e^{-\phi\theta } \right), \nonumber\\
    \theta &= \frac{v_d}{e^{v_d L_{(M)}^d}},\nonumber
\end{align}
where $\mathcal{W}(\cdot)$ is the Lambert-W function, $v_d=\frac{\pi^{d/2}}{\Gamma(d/2+1)}$ is the constant for the $d$-dimensional Lebesgue measure (i.e., $v_d L_{(M)}^d$ is the $d$-dimensional volume of the hyperball with radius $L_{(M)}$), and $\phi$ is the upper bound for $D_t$.
\end{thm}

\begin{IEEEproof}
See the Appendix.
\end{IEEEproof}

Although the expression for $h$ looks complicated, all the terms in \eqref{e:thm} can be easily computed. Particularly, $v_d$ is directly given by the dimensionality $d$, $L_{(M)}$ comes from the training phase, $\phi$ is also found in training, and finally there is a built-in Lambert-W function in popular programming languages such as Python and Matlab. 
Hence, given the training data, $\omega_0$ can be easily computed, and based on Theorem \ref{thm:1}, the threshold $h$ can be chosen to asymptotically achieve the desired false alarm rate.

In the non-asymptotic practical region, where the number of training instances, $N_2$, is finite, the threshold $h$ can still be approximately set using \eqref{e:thm}. The quality of this approximation depends on the amount of training data. With more training data, \eqref{e:thm} will be a better approximation. Also, with more training data, a smaller threshold can be used to achieve the desired false alarm rate $\beta$ as $L_{(M)}, \phi, \theta$ will shrink and in turn $\omega_0$ will grow. Smaller threshold means smaller detection delay, hence more training data increases the performance of the proposed IDS, as expected. If the desired levels of average detection delay and false alarm rate cannot be achieved with the current training set, this will be an indicator for needing more training data. The proposed IDS uses training data to discover nominal data patterns, thus for urban roads with complex patterns, typically more training data is needed compared to a suburban road with simple patterns. 


The selection of other parameters also indirectly affect this fundamental trade-off of quick and accurate detection. Particularly, for bigger/smaller number of neighbors $k$, the proposed IDS becomes more/less robust to noise (i.e., nominal outliers), but at the same time the less/more sensitive to anomalies. In turn, bigger/smaller $k$ result in lower/higher false alarm rate and longer/shorter average detection delay. The parameter $s$ is auxiliary to $k$, and yields similar effects in the algorithm. The significance level $\alpha$ does not play a central role, as opposed to the outlier detection methods, in which $\alpha$ directly controls the essential trade-off between the detection probability (i.e., true positive rate) and false alarm probability (i.e., false positive rate). In the proposed IDS, the affect of the $\alpha$ choice can be compensated by the decision threshold $h$, which is the ultimate parameter that directly controls the balance in detection performance. Hence, in practice, first a typical value, such as $0.05$, is selected for $\alpha$, and then $h$ is chosen to satisfy a desired false alarm rate, as shown in Theorem \ref{thm:1}. 


Detecting an attack is in general not the final task for successful mitigation. Each message from vehicles is needed to be analyzed for general traffic management systems. A powerful IDS should not ignore the data traffic from an anomalous vehicles as a whole; it should identify the anomalous data dimension as soon as possible in order not to allow the attacker to deteriorate the decisions of control center for specific message dimensions. For instance, attacker may only report wrong speed information and the other data contents, e.g., direction and position, may still be valuable for RSU. Another motivation for an in-depth mitigation strategy is to inform other RSUs about what data content is under attack so that they can get prepared beforehand. Without a  mitigation strategy, after the detection, RSU will not only disregard the entire data traffic from anomalous vehicles, but also will not know what specific content of the messages is under attack. 
To this end, we next propose a statistical anomaly localization technique for the proposed IDS to identify the anomalous data dimensions where attacker injects falsified data. 

In (\ref{e:tmin}), detection occurs due to an increase in the decision statistic $s_t$, given by (\ref{e:st}), which is caused by recent positive anomaly evidence(s) $D_t$, given by (\ref{e:Dt}). Moreover, positive $D_t$ happens due to the total distance $L_t$ being greater than the baseline $L_{(M)}$. From (\ref{e:length}), we know that $L_t$ is the sum of $s$ Euclidean distances of data instance $\vx_t$ to its $\{k-s+1,\ldots,k\}$th nearest neighbors. Since, for $\gamma=2$, each Euclidean distance to a neighbor is the sum of squared distances in the $d$ data dimensions, $L_t$ can be written in the following alternative form

\begin{equation}
L_t = \sum_{n=1}^d \ell_t^n ~~\text{where}~~ \ell_t^n = \sum_{j=k-s+1}^k (x_t^n-y_j^n)^2,
\end{equation}
and $x_t^n$ and $y_j^n$ are the $n$th dimensions of the data instance $\vx_t$ and its $j$th nearest neighbor $\vy_j$. Note that $\ell_t^n$ denotes the contribution of dimension $n$ to the total distance $L_t$. Thus, after detection, we can investigate each dimension's contribution to the attack alarm by analyzing some recent $\ell_t^n$ right before the detection time $T$. We determine the number of recent $\ell_t^n$ values contributing to alarm by first identifying the last time instance $q=\max\{t<T: s_t=0\}$ when the decision statistic $s_t$ was zero and then started increasing, which can be seen as an estimate for the attack onset time. Then, the average contribution from each dimension $n$ to the attack alarm is computed as

\begin{equation}
\label{e:ln}
\bar{\ell}^n  =  {(T-q)}^{-1} \sum\limits_{t=q+1}^{T} \ell_t^n,
\end{equation}
where $T$ is the detection time, given by (\ref{e:tmin}). Finally, each dimension $n$ is identified as attacking if its average contribution is sufficiently high, i.e., $\bar{\ell}^n \ge \lambda$. The threshold $\lambda$ controls the trade-off between false positive and true positive rates, as shown in Fig. \ref{f:roc}. It is typically selected to satisfy a constraint on the false positive rate \cite{poor2013introduction}.

The proposed attack detection and localization technique is summarized in Algorithm \ref{alg:ids}.

\begin{algorithm}                      
\caption{Proposed detection \& localization algorithm}          
\label{alg:ids}                           
\begin{algorithmic}[1]                    
\STATE {\emph{Inititalization}:} $s_0 \gets 0$, $t \gets 0$
\STATE \underline{\emph{Training phase:}}
\STATE Partition training set $\cX_{N}$ into $\cX_{N_1}$ and $\cX_{N_2}$
\STATE Compute $L_{i}$ for each $x_i  \: \epsilon \: \cX_{N_1}$  as in \eqref{e:length}
\STATE Find $L_{(M)}$ by selecting the $M$th smallest $L_i$
\STATE \underline{\emph{Test phase:}}
\WHILE{$s_t < h$}
	\STATE $t \gets t+1$
	\STATE Get new data $x_t$ and compute $D_t = (L_t)^d - (L_{(M)})^d$
	\STATE $s_t = \max\{s_{t-1}+D_t,0\}$     
\ENDWHILE
\STATE Attack detected at time $T=t$
\STATE Estimate attack start time as $q=\max\{ t < T: s_t=0\}$
\FOR{$n = 1,\ldots,d$}
	\STATE Compute $\bar{\ell}^n$ as in \eqref{e:ln}
    \IF{$\bar{\ell}^n \ge \lambda$}
        \STATE Declare dimension $n$ as under attack
    \ENDIF
\ENDFOR

\end{algorithmic}
\end{algorithm}

\subsection{Proposed Detection and Mitigation Method for DDoS}
\label{ddos}

A straightforward approach to DDoS detection is through comparing the incoming message rate (i.e., number of messages per unit time) from each vehicle with a threshold. 
Although this method can stop brute-force attacks in which attackers transmit burst of data messages at a high rate, it would not be effective against stealthy DDoS attacks in which multiple attackers transmit messages at a slightly-higher-than-nominal rate synchronously, yet together they overwhelm the RSU. Examples of this new genre of stealthy DDoS attack are recently seen in the Internet, called Mongolian DDoS attack \cite{nexusguard}. 
Admittedly, the straightforward approach that compares the total data rate from all vehicles with a threshold can easily detect any DDoS attack (either brute-force or stealthy); however, it cannot identify the attackers to stop the attack. 

\begin{figure}[tbh]
\centering
\includegraphics[width=.5\textwidth]{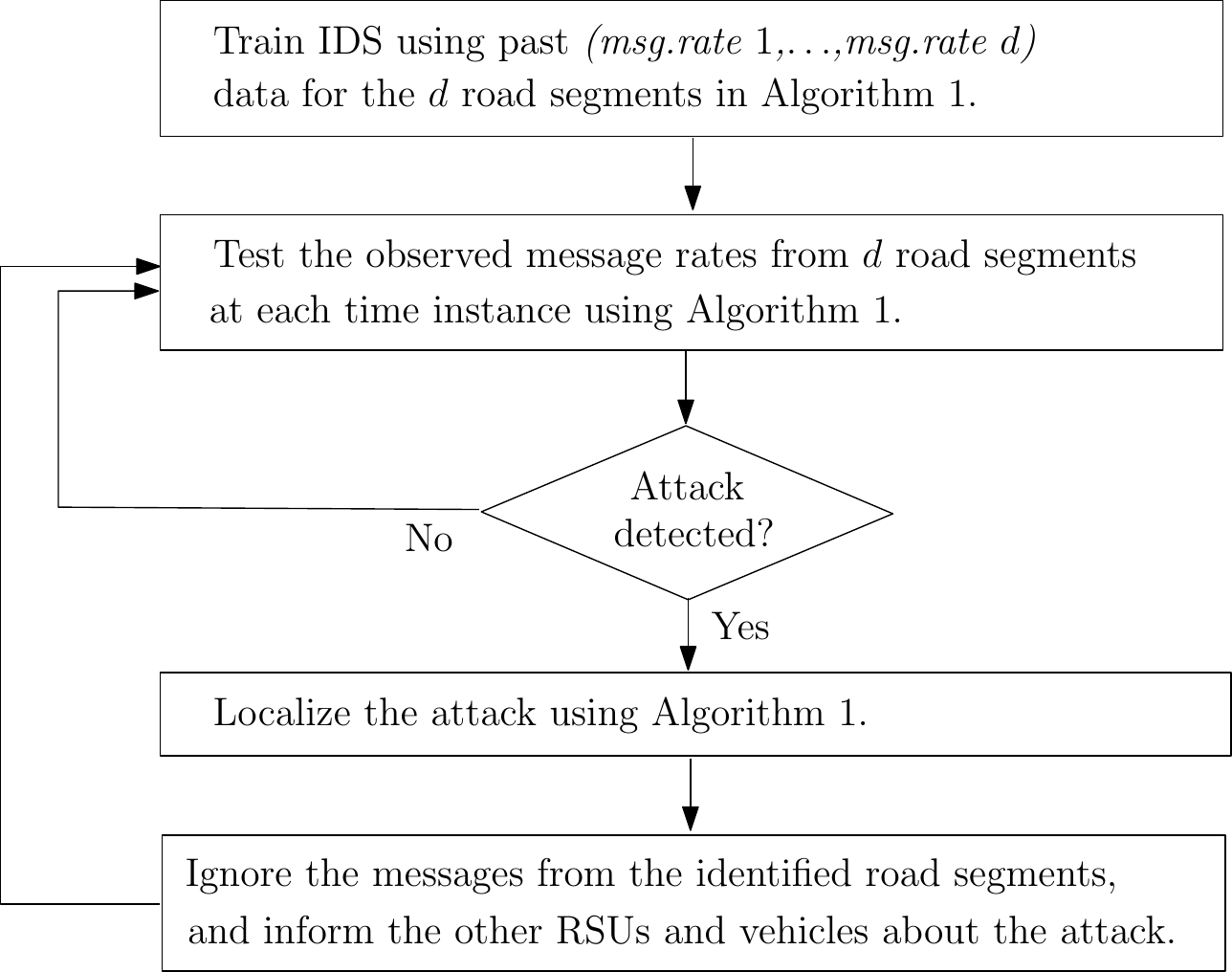}
\vspace{-3mm}
\caption{Flowchart of the proposed IDS for DDoS attacks.}
\label{f:ids-ddos}
\vspace{-1mm}
\end{figure}

We propose a multivariate statistical IDS that jointly monitors vehicles using Algorithm \ref{alg:ids} for detecting and mitigating both stealthy and brute-force DDoS attacks. 
While univariate methods will fail to detect the malicious vehicles' close-to-nominal data rates in stealthy DDoS, our proposed multivariate approach can easily detect them as simultaneous small increases from multiple vehicles cause a considerable increase in $k$NN distance with respect to the nominal training data. 

Since the number of vehicles in the range of an RSU varies over time, for joint monitoring of data traffic, we consider total message rates in a number of predetermined road segments as the input data to the proposed algorithm, $\vx_t=[x_t^1,\ldots,x_t^d]$ where $x_t^n$ is the message rate from road segment $n$ at time interval $t$ (see Fig. \ref{f:attack}). 
Detecting DDoS attack is not the only goal of an effective IDS; it should also mitigate the detected attack to protect the VANET. 
Especially for stealthy DDoS attacks, identifying the attacking nodes is a challenging task that must follow detection. Otherwise, after the detection, RSU will either disregard the entire data traffic or wait with no further action until the excessive incoming data paralyzes it. In either case, the DDoS attack would be successful in making the RSU service unavailable. Our proposed attack mitigation strategy will be activated after DDoS attack is detected to identify the anomalous road segment whose data will be blocked. Since the proposed algorithm is sequential in nature, it can dynamically detect and mitigate DDoS attacks due to moving vehicles in real-time. The proposed IDS for mitigating DDoS attacks is summarized in Fig. \ref{f:ids-ddos}. 

\section{Performance Evaluation}
\label{s:sim}

\begin{figure}[t]
\begin{subfigure}
\centering
\includegraphics[width=0.45\textwidth]{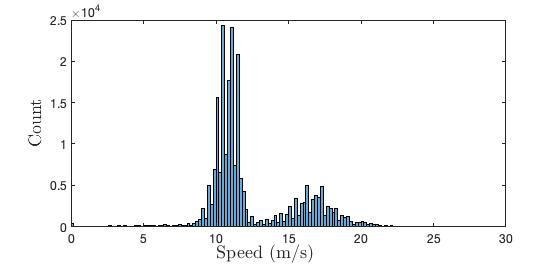}
\end{subfigure}
\begin{subfigure}
\centering
\includegraphics[width=0.45\textwidth]{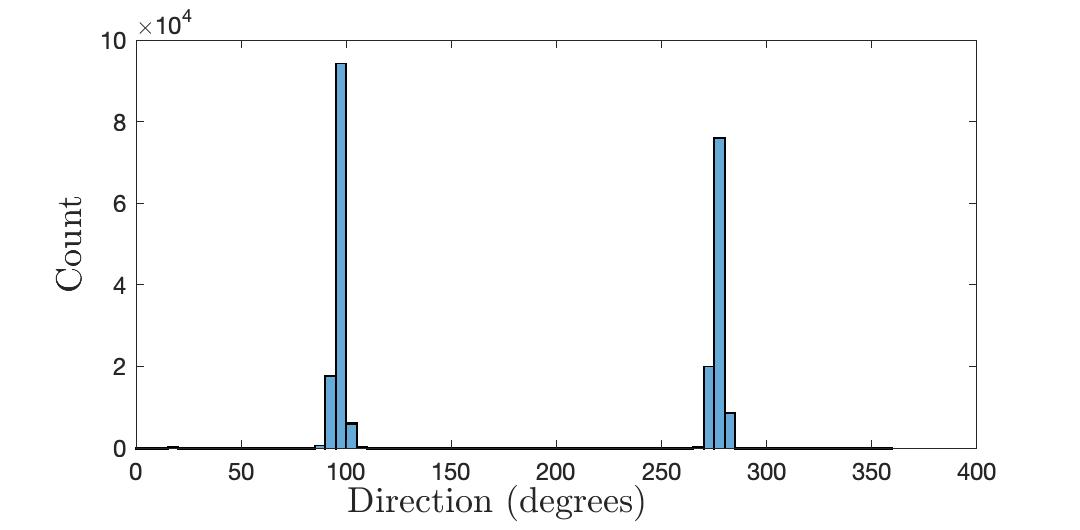}
\end{subfigure}
\begin{subfigure}
\centering
\includegraphics[width=0.45\textwidth]{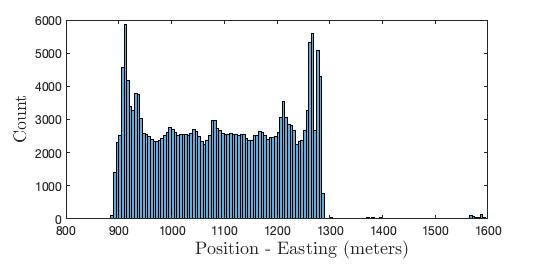}
\end{subfigure}
\caption{The heterogeneous probability distributions of message contents in the Warrigal dataset. Histograms are obtained from the training set.}
\label{f:dataset2}
\vspace{-2mm}
\end{figure}

\begin{figure*}[ht!]
\subfigure[Anomaly implemented in speed]
{
\centering
\includegraphics[width=0.33\textwidth]{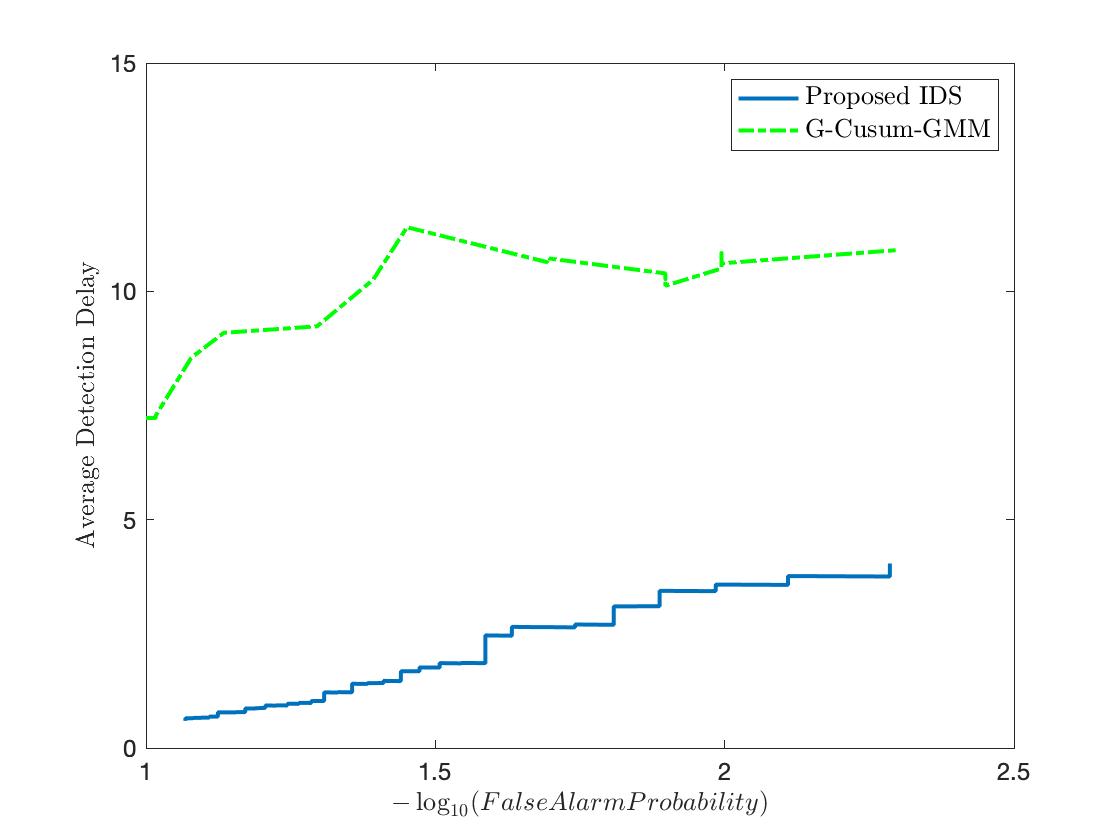}
\label{f:odit1}
}
\subfigure[Anomaly implemented in position (easting)]
{
\centering
\includegraphics[width=0.33\textwidth]{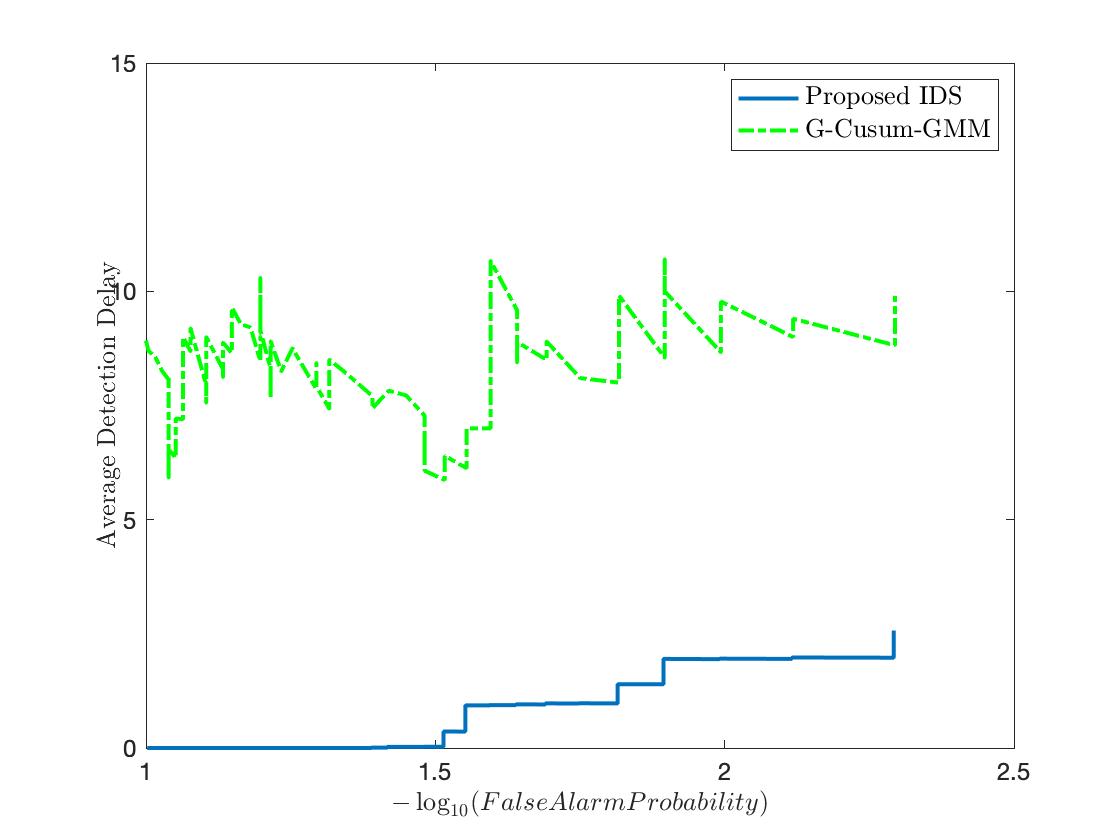}
\label{f:odit2}
}
\subfigure[Anomaly implemented in direction]
{
\centering
\includegraphics[width=0.33\textwidth]{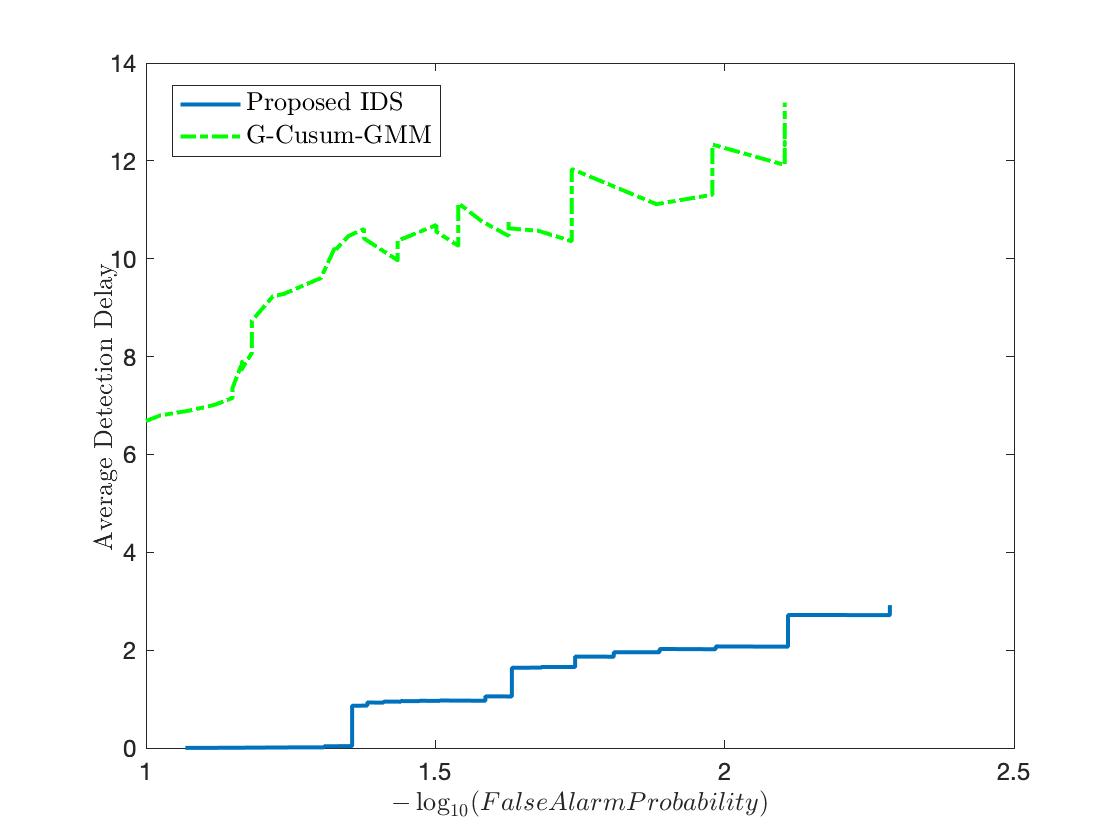}
\label{f:odit3}
}
\caption{Comparison in terms of quick and accurate detection under different FDI attack scenarios between the proposed detector and an idealized version of the state-of-the-art sequential detector (G-CUSUM) which exactly knows the attack magnitudes.}
\label{f:resultsFDI}
\end{figure*}

\begin{figure*}[t]
\subfigure[Detection rate vs. Number of attackers]
{
\centering
\includegraphics[width=0.5\textwidth]{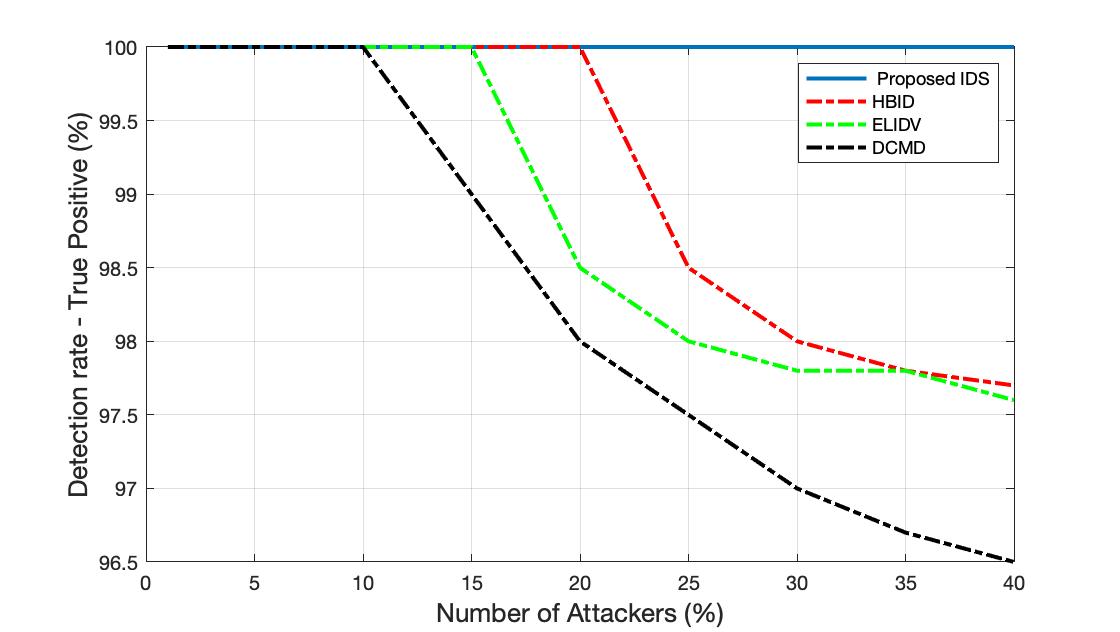}
\label{f:comp1}
}
\subfigure[False positive rate vs. Number of attackers]
{
\centering
\includegraphics[width=0.5\textwidth]{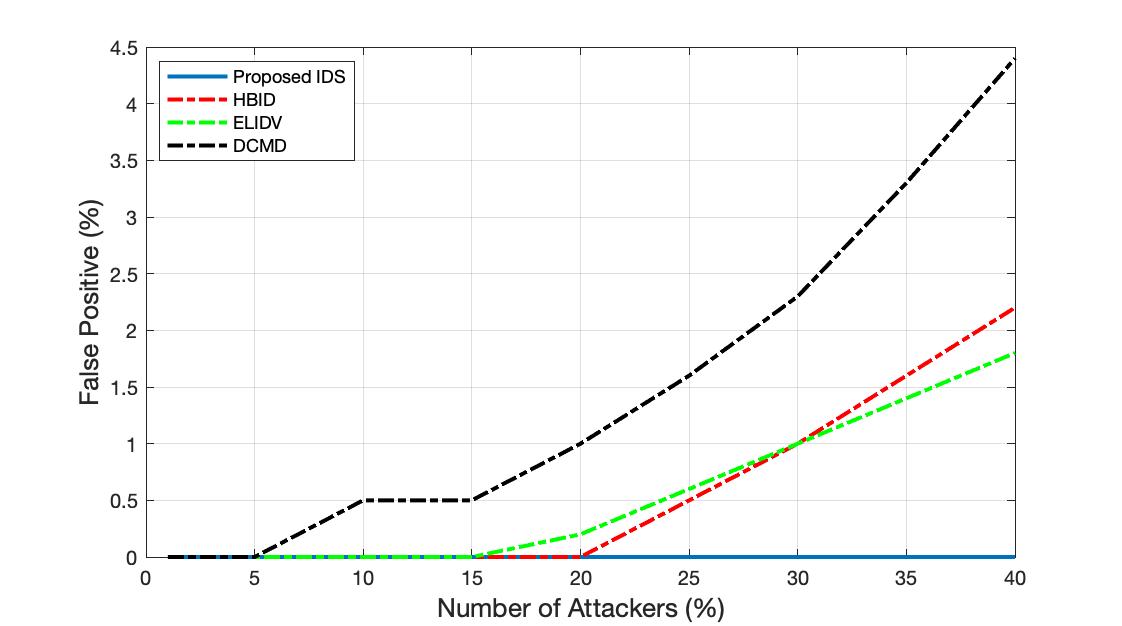}
\label{f:comp2}
}
\caption{Comparison considering FDI attack in speed values between the proposed detector and several voting-based detectors from the literature.}
\label{f:comp}
\end{figure*}

In this section, we evaluate the performance of the proposed IDSs using a real dataset for FDI attack and simulated data for DDoS attack. 

\subsection{Detection Results for FDI Attack}
\subsubsection{Experiment Setup}

We use the Warrigal dataset, collected by the University of Sydney in an industrial area over a period of three months with 1 Hz resolution \cite{ward2014warrigal}. Each message in the dataset consists of position, speed and direction information. Position information is given in three dimensions as easting, northing and altitude in meters. Speed and direction values are provided in meter/second and degrees, respectively. The histograms of training data for position (easting), speed and direction are shown in Fig. \ref{f:dataset2}. Due to the heterogeneity, it is not tractable to estimate the joint distribution for parametric methods. Since our IDS runs at an RSU, we consider only a portion of available data which is collected from a few km road range, where RSU is assumed to be located at the center. 

In order to generate the FDI attack scenarios, we separately injected anomalous data to each dimension generated from uniform distribution. In each scenario, anomalous data is injected into one of the data dimensions of a randomly selected set of vehicles (i.e., attacking vehicles). 
For instance, attacking vehicle only falsifies its position while broadcasting to the VANET. Falsifying multiple message dimensions is also possible, but since detecting such case would be easier, we only considered attacks on single data content in our experiments. Based on the nominal rates in Warrigal dataset, anomaly rates for position and direction are selected as $30\%$ and $40\%$ of the nominal values, respectively. These anomaly rates reflect the attackers' objective of disrupting integrity with wrong information while remaining undetected as much as possible. 
Similarly in speed, after anomaly injection, the falsified values of attacking vehicles go up to 22 m/s (50 mph), which is still in the nominal range of training data as shown in Fig. \ref{f:dataset2}. 
Note that even slight falsification in speed, position, and direction from multiple vehicles can cause trouble in RSU's traffic management, as well as other vehicles' decisions. 
To create a challenging scenario for detection and localization, in each test, anomaly is inserted in one of the message dimensions for only 20 seconds. 


\subsubsection{Results}

\begin{figure}[t]
\centering
\includegraphics[width=.5\textwidth]{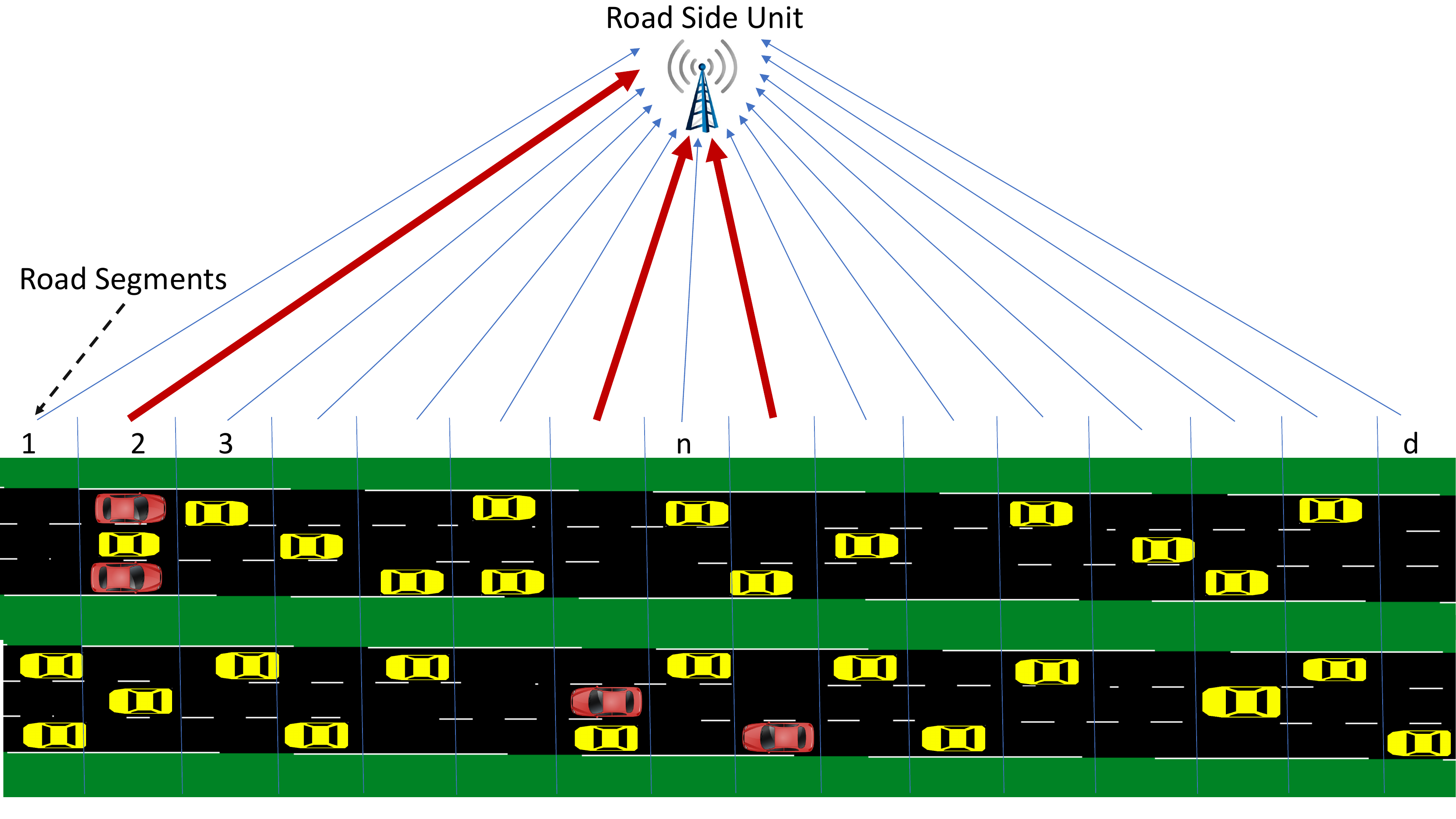}
\caption{DDoS attack model where red cars are attackers and thick red lines denote the increased data rates.}
\label{f:attack}
\end{figure}

We compare the performance of proposed IDS on FDI attacks with state-of-the-art sequential and voting-based methods in the literature.

We start with comparing the quick and accurate detection performance of the proposed IDS with an idealized version of the state-of-the sequential detector, G-CUSUM, which fits a probability distribution to nominal data and somehow exactly knows the attack magnitudes in the anomalous data. In practice, it is not tractable for G-CUSUM to know the actual attack magnitudes. Since the data distribution for each dimension in Fig. \ref{f:dataset2} is close to a mixture of two Gaussian distributions, G-CUSUM assumes a Gaussian Mixture Model (GMM) type of probability distributions for both nominal and anomaly data. Three FDI attack scenarios are investigated for anomalies in the speed, position, and direction data, whose results are given in Fig. \ref{f:odit1}, \ref{f:odit2}, and \ref{f:odit3}, respectively. Fig \ref{f:odit2} shows the results for the case when anomaly is in the easting dimension of position. Similar results are observed with the northing dimension of position as well. The data-driven nature of the proposed IDS enables much quicker detection while satisfying the same false alarm rates compared to G-CUSUM. The proposed IDS learns the nominal baseline from data and detects the deviations from this baseline, whereas G-CUSUM suffers from the mismatches between the assumed and actual probability distributions for the nominal and anomalous data. 

Since voting-based IDS is a popular choice in the literature, we next compare the proposed IDS with a number of voting-based IDSs, namely HBID \cite{zaidi2016host}, ELIDV \cite{sedjelmaci2014efficient}, and DCMD \cite{ruj2011data}. These systems run on each vehicle where each received message content is examined with a voting scheme. A main performance difference between such models and the proposed IDS is that while the detection accuracy for these systems decreases with increasing number of anomalous vehicles due to the inherent rules of voting, the proposed IDS is not affected since each vehicle is monitored at the RSU. This fact is illustrated in Fig. \ref{f:comp1} and \ref{f:comp2} in terms of true positive rate and false positive rate, respectively, considering the anomalous speed scenario. The proposed detector achieves $100\%$ detection (true positive rate) and $0\%$ false alarm (false positive rate) within 12 seconds in the considered FDI attacks regardless of the number of attackers. Whereas, the performance of voting schemes quickly degrades after the percentage of attackers in the entire vehicle population reaches a certain threshold. 

\subsection{Detection Results for DDoS Attack}
\subsubsection{Experiment Setup}

In this section, we evaluate the performance of proposed IDS for stealthy DDoS attacks targeting the availability of VANET communications. As shown in Fig. \ref{f:attack}, we split an RSU range into a number of road segments with equal length. Total number of received messages from a road segment in a time interval (message rate) gives the data dimension from that road segment, which mainly depends on the number of vehicles and speed of vehicles. For example, if traffic flow decreases, leading to a rise in the number of vehicles on the road, the message rate from the road segments increases.

\begin{figure}[b]
\centering
\includegraphics[width=.45\textwidth]{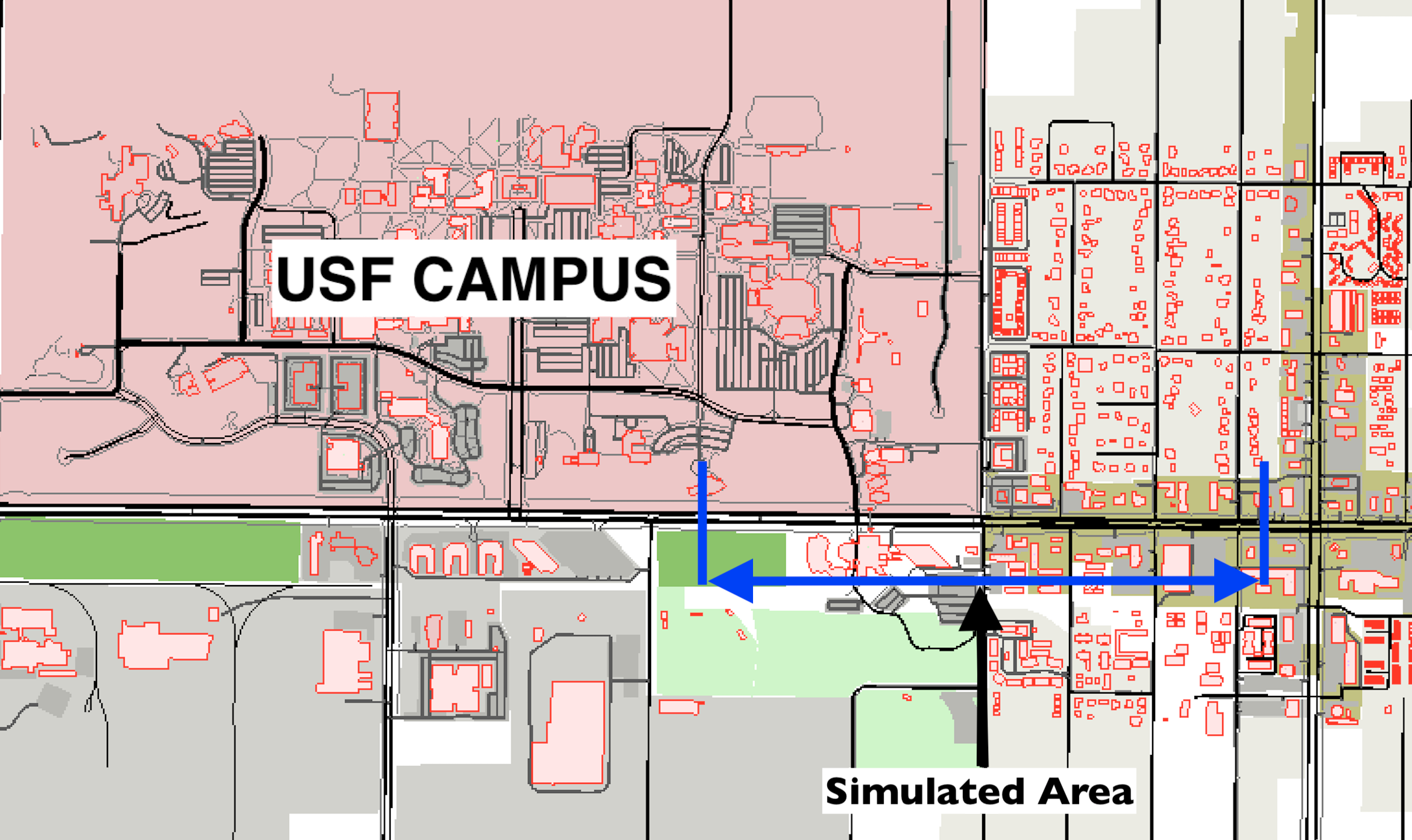}
\caption{Simulation map showing Fowler Ave.}
\label{f:map}
\end{figure}

For the simulation study, we use three frameworks together: OMNET++ \cite{varga2008overview}, SUMO \cite{behrisch2011sumo}, and Veins \cite{sommer2011bidirectionally}. OMNET++, which is a general network simulator, creates a VANET environment. Simulation of Urban Mobility (SUMO) and Veins are the two supportive frameworks, where SUMO provides a mobility model for VANET and Veins creates an interface between SUMO and OMNET++. While vehicles are moving on the roads in SUMO, they are identified as a mobile node in OMNET++ by the help of Veins. We based our simulations on the IEEE 802.11p vehicular communication protocol \cite{jiang2008ieee}, but since our model does not specify any communication protocol, our DDoS detection algorithm can be used with other protocols as well.

We simulate a realistic scenario with SUMO by using a real road map, which is a small section of Fowler Ave. next to the University of South Florida (USF) campus in Tampa, Florida (See Fig. \ref{f:map}). Selected road section is partitioned into 20 segments with 50 meters width for each road segment. In order to have a realistic dataset, there is no restriction on vehicular movements; all vehicles follow their randomly generated routes, i.e., they can join or leave the main road at any intersection. Average number of vehicles in the simulation area is 250. 

\begin{table}[t]
\centering
\caption{SIMULATION PARAMETERS}
\begin{tabular}{ |p{3.5cm}|p{2.8cm}|  }
 \hline
 Simulation Area& 9000 x 5000 $m^2$ \\ \hline
Simulation Time (Each Trial) &	200s\\ \hline
Number of Trials&		600\\ \hline
Average Number of Vehicle&		250\\ \hline
Traffic Generation &Random\\ \hline
Route Generation & Random \\ \hline
Network Protocol & IEEE 802.11p\\ \hline
Beacon Rate&	1s\\ \hline
Network Interface&	OMNET++\\ \hline
Network Mobility Framework&	Veins \\ \hline
Traffic Generator&	SUMO\\ \hline
Map& Fowler Av. Tampa, FL \\ \hline
\end{tabular}
\vspace{3mm}
\label{table:1}
\end{table}

With the given simulation parameters in Table \ref{table:1}, 4 hours of traffic is observed for learning the training baseline and 33.3 hours of traffic is observed for the test purposes. After saving all the log files, data rates for each road segment are calculated on MATLAB, and 600 test trials of 200-second duration are obtained. We generated anomaly data in MATLAB from uniform distribution for two different DDoS attack scenarios. We consider 0.3 times mean increase for the first scenario, and 1.5 times mean increase for the second scenario with respect to the corresponding nominal baseline. Anomalies are inserted on top of the nominal data in 2 of the 20 road segments from 181st second to 200th second. 

\subsubsection{ Results}

\begin{figure}[t]
\begin{subfigure}
\centering
\includegraphics[width=0.45\textwidth]{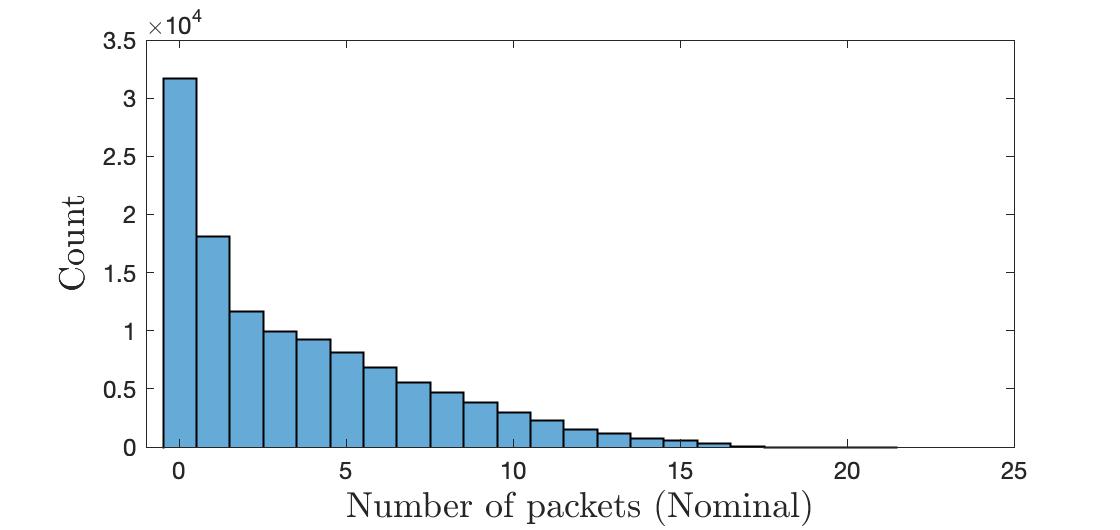}
\end{subfigure}
\begin{subfigure}
\centering
\includegraphics[width=0.45\textwidth]{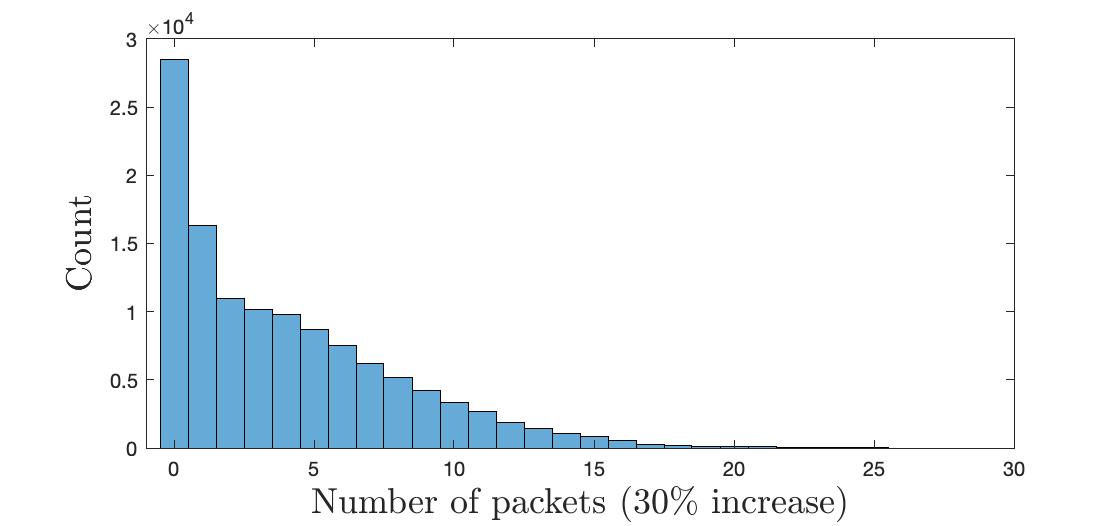}
\end{subfigure}
\begin{subfigure}
\centering
\includegraphics[width=0.45\textwidth]{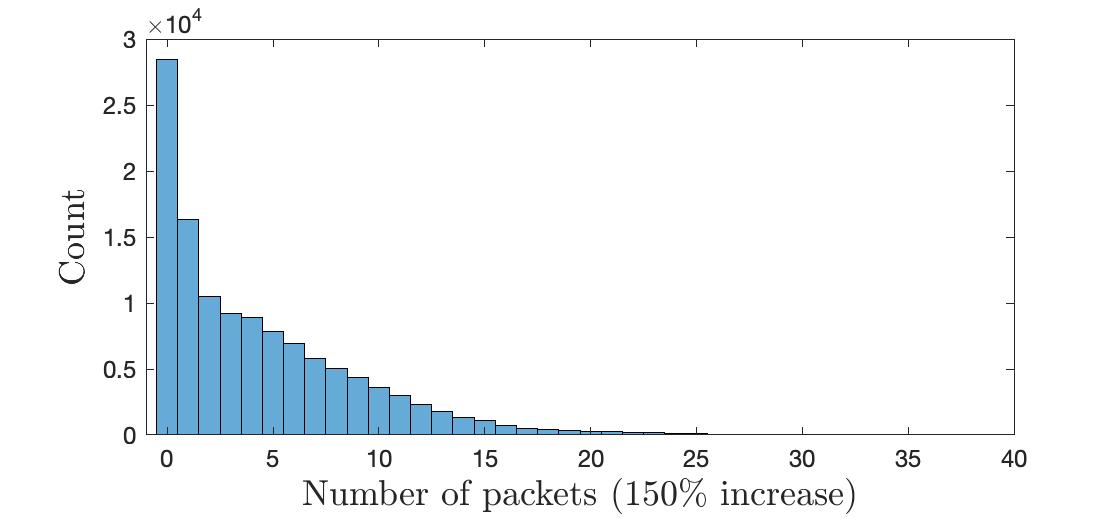}
\end{subfigure}
\caption{Histogram of number of packets for a road segment. First histogram represents the distribution of nominal data, whereas second and third represent attack cases with an average increase that is 0.3 and 1.5 times the baseline, respectively. Nominal and attack distributions are close to negative binomial distribution with extended tails under attacks.}
\label{f:dataset}
\end{figure}

\begin{figure}[t]
\begin{subfigure}
\centering
\includegraphics[width=0.45\textwidth]{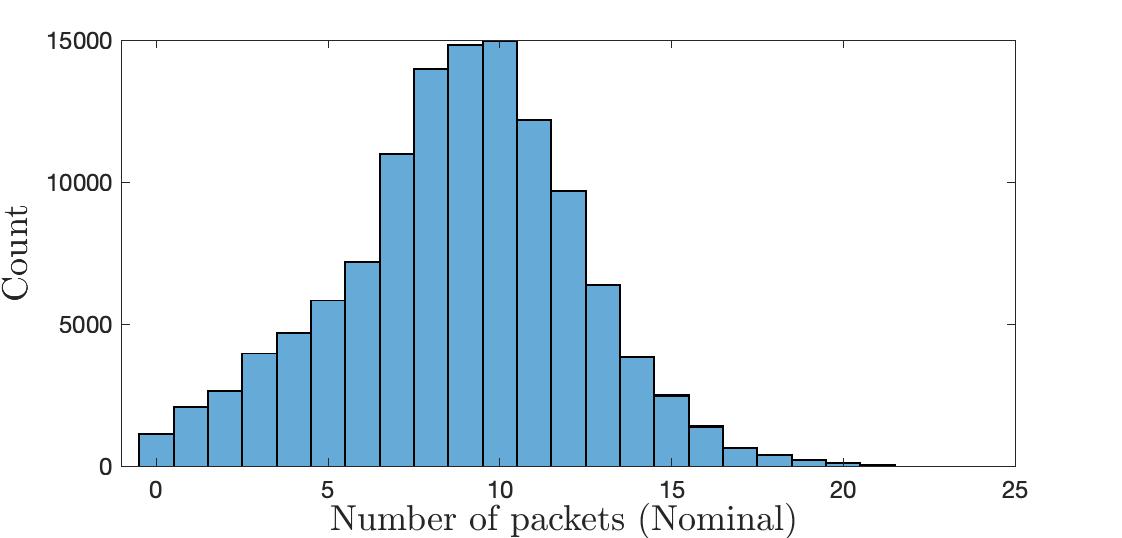}
\end{subfigure}
\begin{subfigure}
\centering
\includegraphics[width=0.45\textwidth]{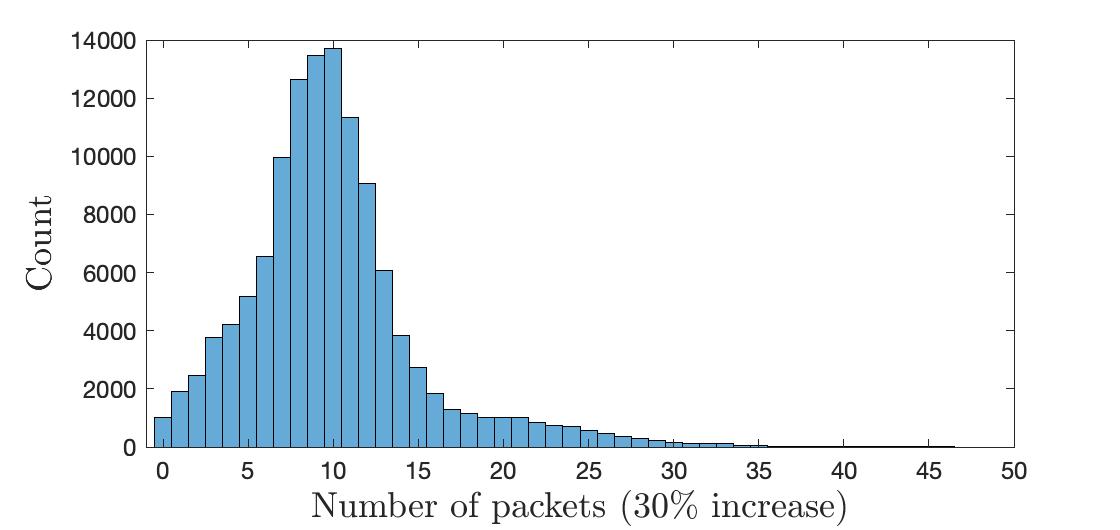}
\end{subfigure}
\begin{subfigure}
\centering
\includegraphics[width=0.45\textwidth]{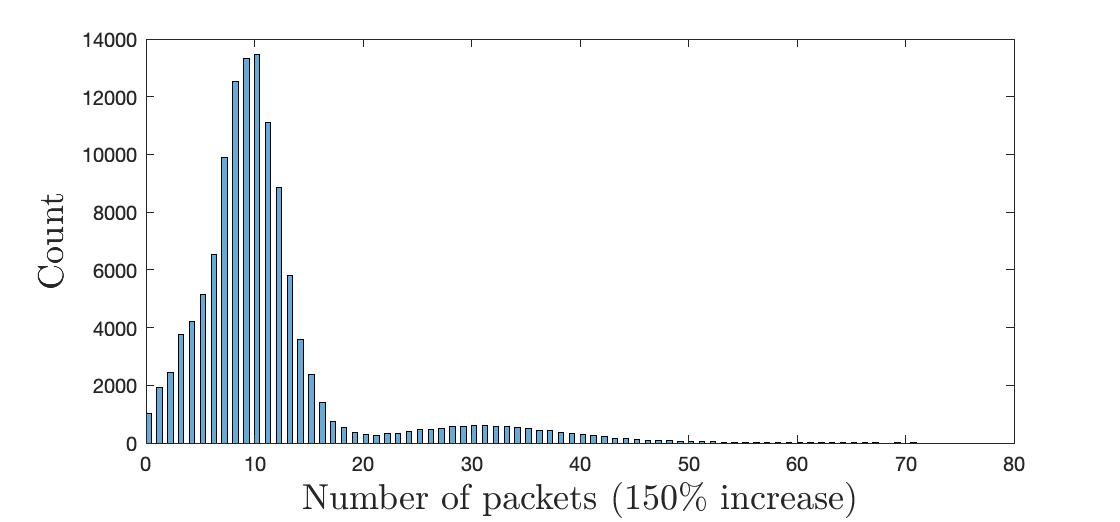}
\end{subfigure}
\caption{Histogram of number of packets for a road segment. First histogram represents the distribution of nominal data, whereas second and third represent attack cases an average increase that is 0.3 and 1.5 times the baseline, respectively. Nominal and attack distributions are close to normal distribution with extended tails under attacks.}
\label{f:dataset1}
\end{figure}

We compare the proposed IDS with the state-of-the-art sequential G-CUSUM detector, e.g., \cite{guo2010forensic}, and the data filtering method, e.g., \cite{verma2013prevention}. G-CUSUM assumes a probability distribution for nominal and anomalous data, whereas the data filtering approach looks for an increase in the total data rate received by RSU without performing any statistical analysis. In Fig. \ref{f:dataset} and Fig. \ref{f:dataset1}, it is seen that the observations from the two road segments to which anomaly is added follow different distributions. While the distribution of one road segment is similar to negative binomial (Fig. \ref{f:dataset}), which is indeed a Poisson distribution with conjugate prior (i.e., Gamma distribution) on the rate parameter, the distribution of other road segment is similar to Gaussian (Fig. \ref{f:dataset1}). Hence, we examine two idealized versions of G-CUSUM which fit negative binomial and Gaussian distributions for each road segment, and somehow exactly know the attack magnitudes of 30\% and 150\%. 
 
\begin{figure}[t]
\centering
\includegraphics[width=0.5\textwidth]{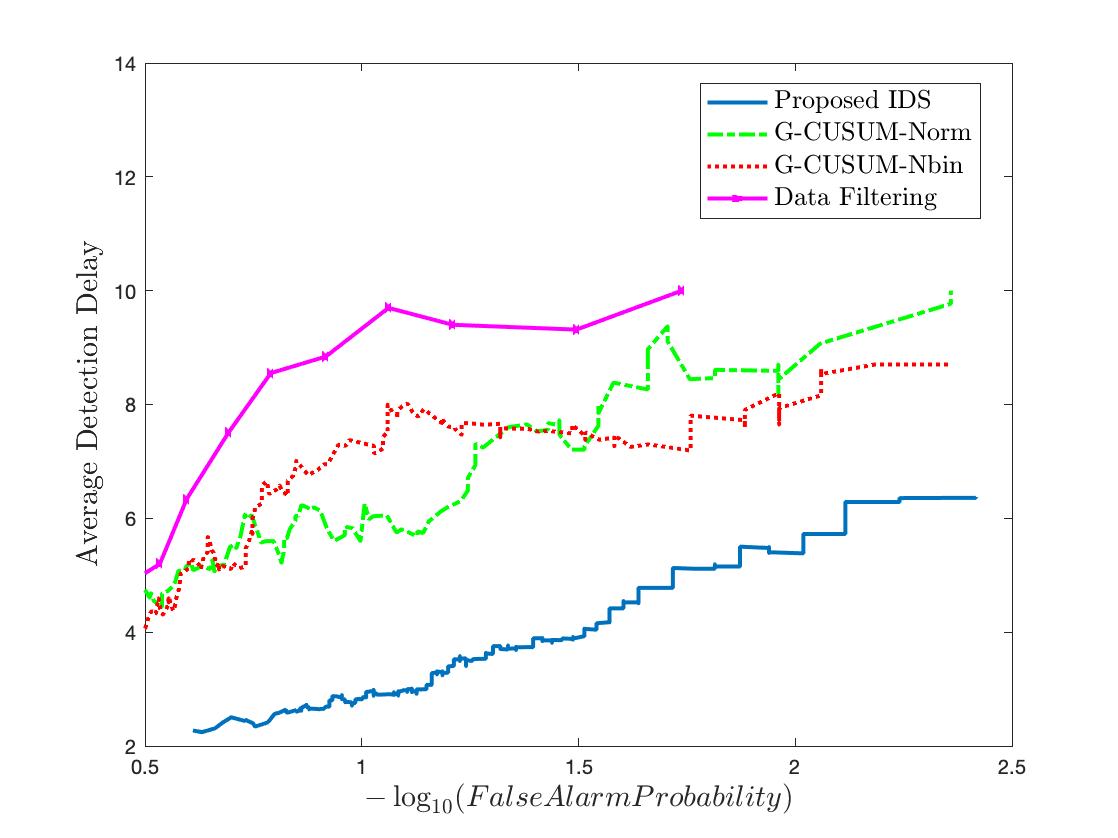}
\caption{Comparison in terms of quick and accurate detection for an average DDoS attack magnitude of 0.3 times the nominal mean data rate between the proposed method, two idealized G-CUSUM variants which exactly know the attack magnitude, and the data filtering method.}
\label{f:odit4}
\end{figure}

\begin{figure}[t]
\centering
\includegraphics[width=0.5\textwidth]{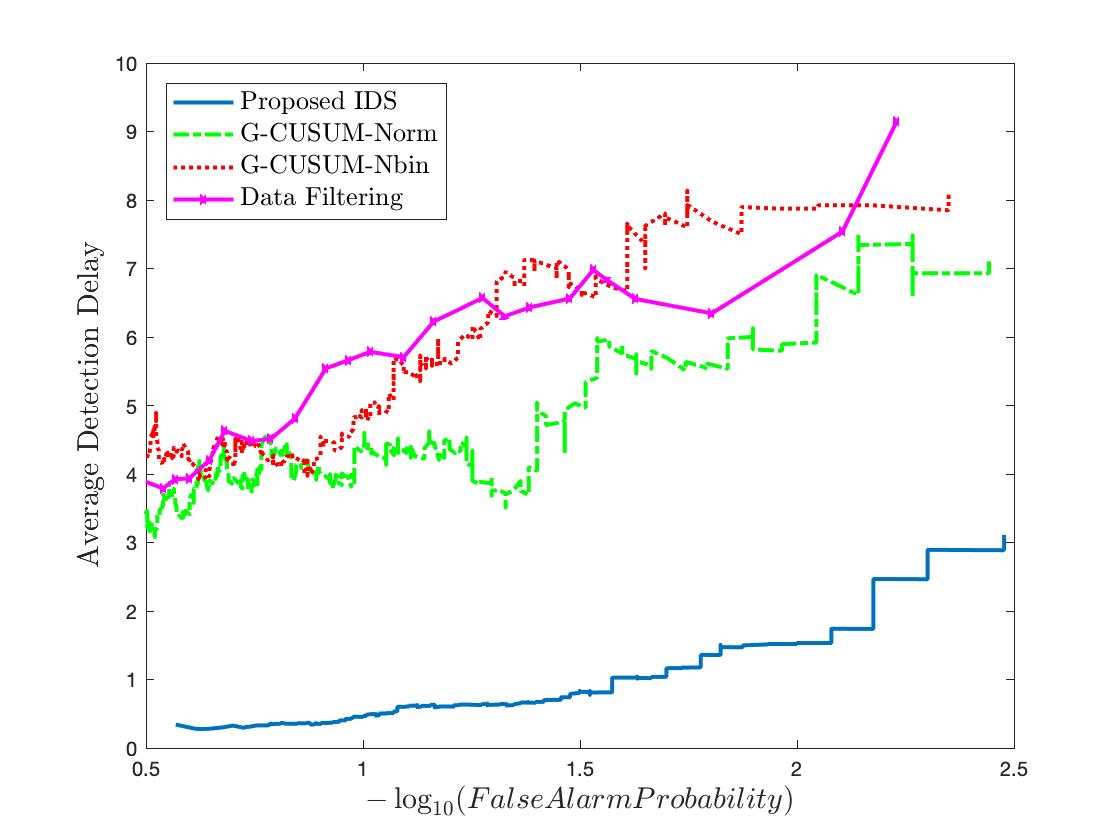}
\caption{Comparison in terms of quick and accurate detection for an average DDoS attack magnitude of 1.5 times the nominal mean data rate between the proposed method, two idealized G-CUSUM variants which exactly know the attack magnitude, and the data filtering method.}
\label{f:odit5}
\end{figure}

For both attack scenarios with 30\% and 150\% average mean increase from the nominal mean rate, Fig. \ref{f:odit4} and Fig. \ref{f:odit5} show that the proposed IDS outperforms the G-CUSUM approach and the data filtering approach in terms of quick accurate detection. In particular, the proposed IDS achieves much smaller average detection delay while satisfying the same false alarm rates (e.g., for $0.01$ false alarm rate, approximately $1/2$ times and $1/5$ times in Figs. \ref{f:odit4} and \ref{f:odit5}). Moreover, the G-CUSUM and data filtering approaches have certain practical disadvantages compared to the proposed IDS. The data filtering method can only detect such low-rate stealthy attacks by monitoring the total number of packets received by the RSU since the individual data rates from road segments still appear to be harmless to the network. As a result, it is not tractable for the data filtering method to localize and mitigate the attack. For G-CUSUM, indeed there is no way to exactly know the actual attack magnitudes. In practice, a number of parallel tests with different assumptions for the attack magnitude can be applied, however even for the best test that alarms first, the mismatch between the assumed anomaly distribution and the actual distribution would cause significant performance degradation. As shown in Figs. \ref{f:odit4} and \ref{f:odit5}, even the ideal G-CUSUM which exactly knows the attack magnitude suffers from the deviations of the observed data from the assumed probability distributions. Furthermore, G-CUSUM inevitably follows a univariate approach by assuming independence between road segments \cite{mei2010efficient} since it does not know which road segments will include anomaly. The multivariate nature of the proposed detector also facilitates its superior performance.

\begin{figure*}
\subfigure[Identification of anomalous vehicles in FDI attack to speed data.]
{\centering
\includegraphics[width=0.33\textwidth]{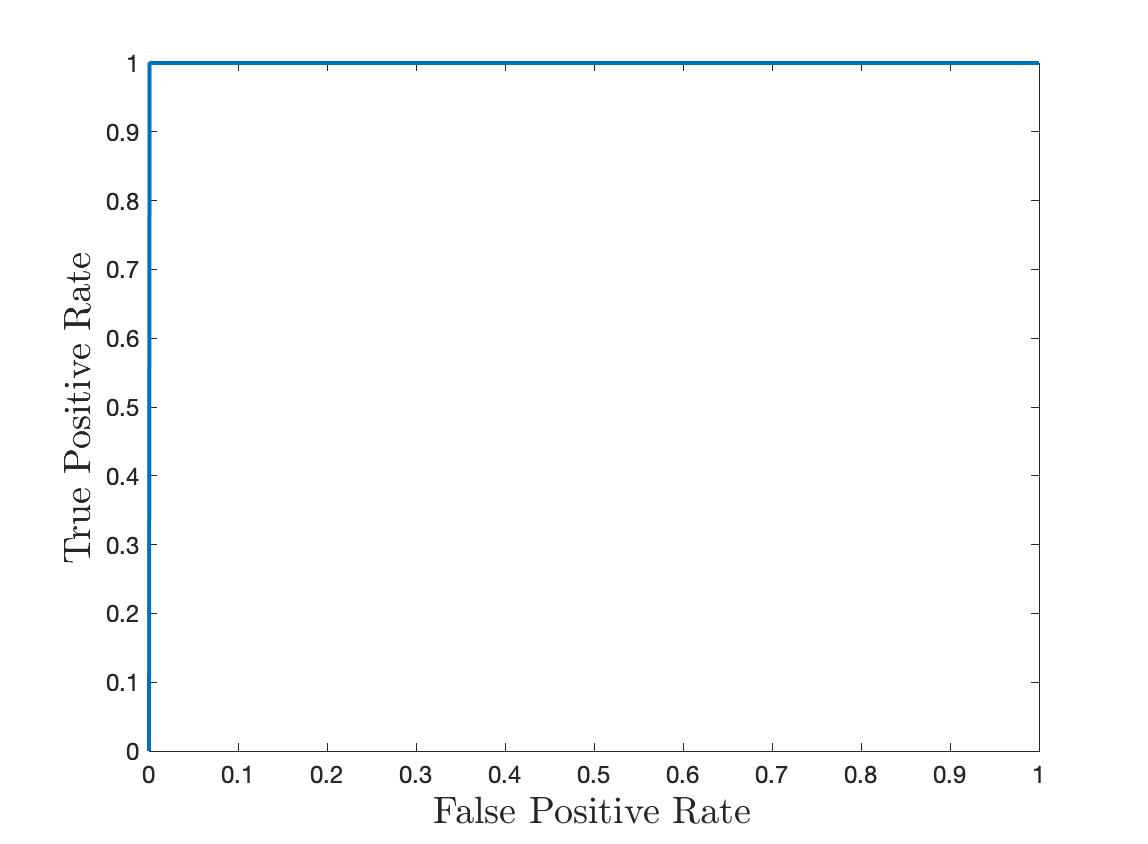}
\label{f:roc1}
}
\subfigure[Identification of anomalous data in FDI attack to speed data.]
{\centering
\includegraphics[width=0.33\textwidth]{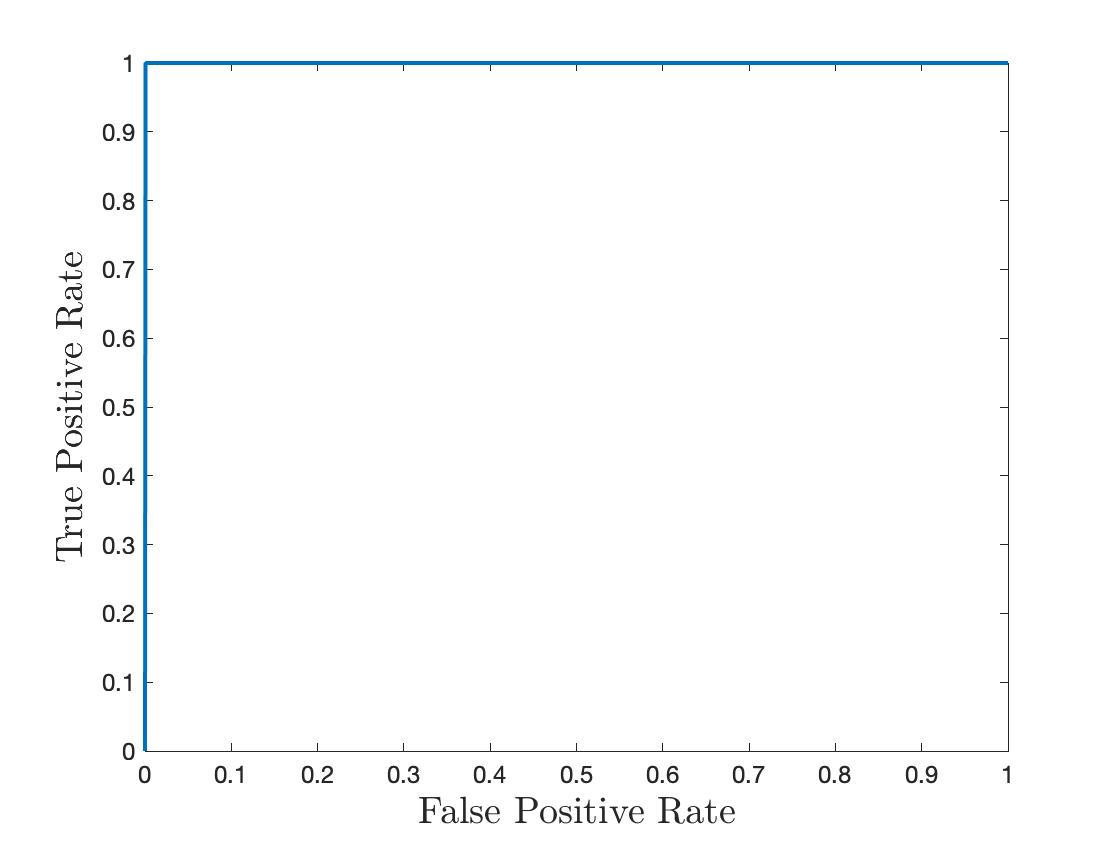}
\label{f:roc2}
}
\subfigure[Identification of anomalous road segments in DDoS attack.]
{\centering
\includegraphics[width=0.33\textwidth]{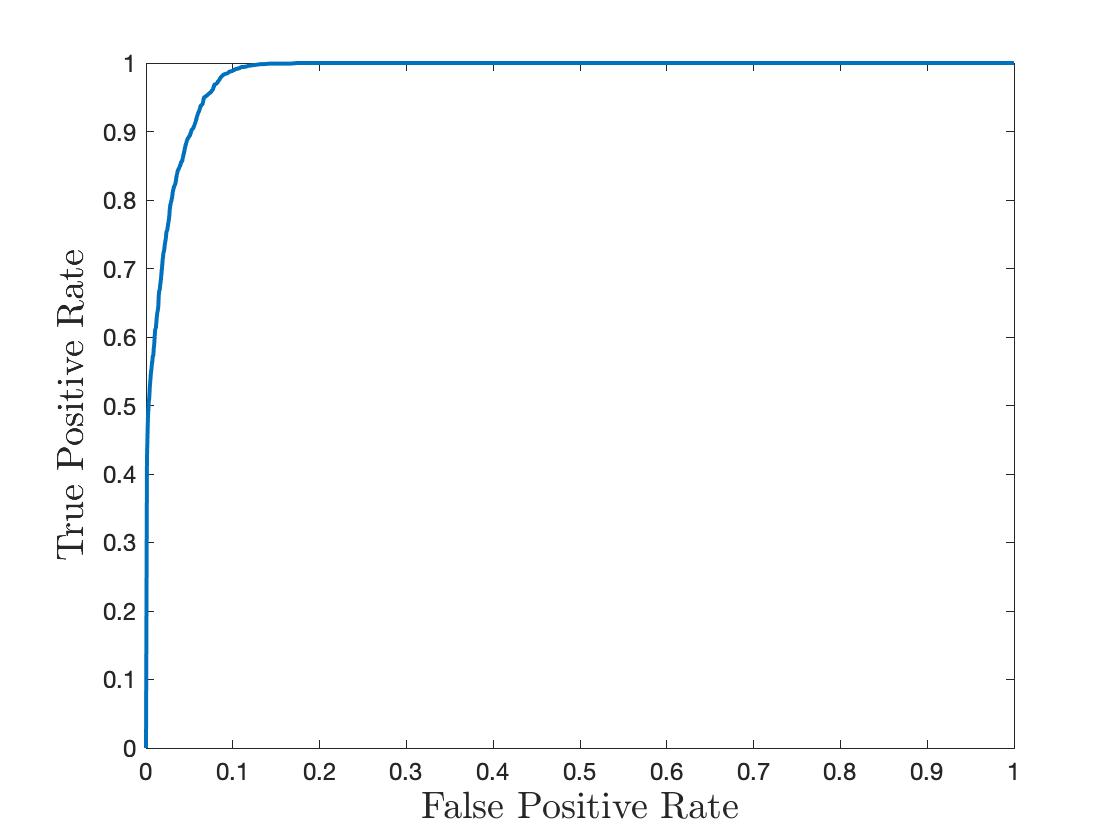}
\label{f:roc3}
}
\caption{ROC curves for the proposed IDS's anomaly localization performance.}
\label{f:roc}
\end{figure*}

\subsection{Localization Results}

We next evaluate the attack localization performance of the proposed IDS using the receiver operating characteristic (ROC) curves, which presents the achieved true positive rates while the algorithm satisfies different levels of false positive constraints. Firstly, we consider the identification of attacking vehicles in the FDI attack scenario (anomalous speed case). Since in this case the proposed detector is applied to each vehicle and the messages include the vehicle ID, there is no need for a separate vehicle identification mechanism after detection. Specifically, once the proposed IDS alarms for a vehicle, this vehicle is automatically identified as attacking. In the anomalous speed scenario, by selecting the detection threshold as $h=2$, in all test trials, the proposed IDS achieves zero false alarm for non-attacking vehicles and 100\% correct detection of attacking vehicles with a maximum delay of 12 seconds (Fig. \ref{f:roc1}).  We next consider the identification of anomalous data dimension using the localization strategy given in (6) and summarized by Algorithm 1. Fig. \ref{f:roc2} displays the perfect detection of the anomalous speed data while satisfying zero false alarm in all test trials. 

Finally, the identification of road segments in the DDoS attack scenario using Algorithm 1 is considered. As demonstrated by Fig. \ref{f:roc3}, the proposed IDS successfully identifies the anomalous road segments with a high correct detection rate (e.g., 94\%) while satisfying a small false alarm rate (e.g., 5\%).

\section{Conclusion}
\label{s:conc}

We proposed a statistical nonparametric intrusion detection system (IDS) for online detection of false data injection (FDI) attacks and distributed denial-of-service (DDoS) attacks. The proposed system runs at roadside unit (RSU) monitoring the broadcasted messages from the vehicles in its range. To be specific, in the FDI attack case, we considered the (ID, speed, position, direction) message format; however, the proposed IDS is based on a generic anomaly detection algorithm, and thus easily extends to other data types. Similarly, the IDS proposed for DDoS attacks is applicable to any data type and communication protocol as it monitors the data rates (i.e., number of packets in unit time) from a number of road segments. An attack localization procedure was also proposed to follow up on an alarm raised by the detection procedure. As the final stage in attack mitigation, RSU drops the identified messages from identified vehicles for FDI attacks and from identified road segments for DDoS attacks. The detection and localization performances of the proposed IDS are evaluated in the FDI and stealthy DDoS cases using a real traffic dataset, called the Warrigal dataset, and state-of-the-art traffic simulators, respectively. To the best of our knowledge, this work is the first to use a real dataset in VANET cybersecurity. Experiment results demonstrated the superior performance of the proposed IDS in terms of quick and accurate detection and localization compared to state-of-the-art voting schemes, parametric sequential change detection algorithm, and the data filtering method. As a future work, we plan to extend the proposed IDS to other attack types.

\bibliographystyle{IEEEtran}
\bibliography{References.bib}

\begin{IEEEbiography}
[{\includegraphics[width=1in,height=1.4in,clip,keepaspectratio]{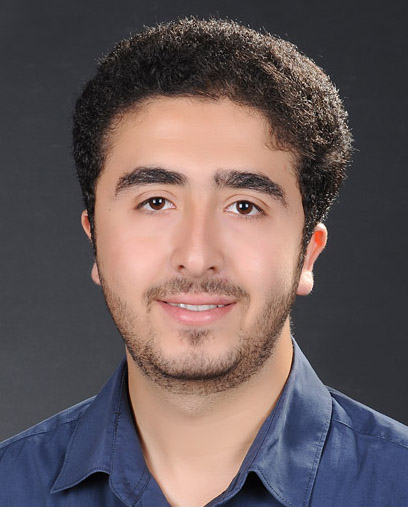}}]{Ammar Haydari}
received the B.Sc. degree in Electronic Engineering from Uludag University, Bursa, Turkey, in 2014 and M.S. degree in Electrical Engineering from University of south Florida, Tampa, FL, in 2019. He is currently a Ph.D. student at the Department of Electrical and Computer Engineering at the University of California, Davis. His research interests include intelligent transportation systems, cyber-security and machine learning. 
\end{IEEEbiography}

\begin{IEEEbiography}
[{\includegraphics[width=1in,height=1.4in,clip,keepaspectratio]{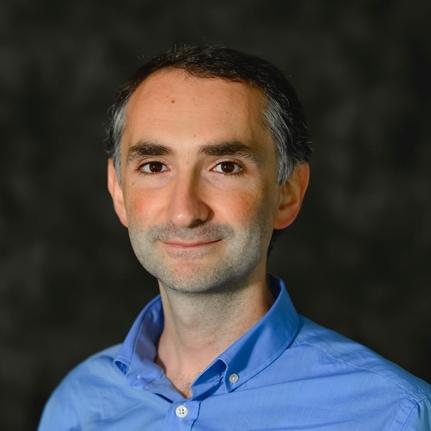}}]{Yasin Y{\i}lmaz}
(S'11-M'14) received the Ph.D. degree in Electrical Engineering from Columbia University, New York, NY, in 2014. He is currently an Assistant Professor of Electrical Engineering at the University of South Florida, Tampa. He received the Collaborative Research Award from Columbia University in 2015. His research interests include statistical signal processing, machine learning, and their applications to cybersecurity, cyber-physical systems, IoT networks, transportation systems, energy systems, and communication systems. 
\end{IEEEbiography}

\end{document}


\nolinenumbers
\twocolumn[
\icmltitle{Proof of Theorem 1}
]

The proof of Theorem 1 is based on an upper bound, $FAR \le e^{-\omega_0 h}$, for the false alarm rate of CUSUM-like algorithms with independent increments \cite{basseville1993detection}, such as the proposed IDS. Here, $\omega_0 \geq 0$ is the solution to $E[e^{\omega_0 D_t}] = 1$. Given a false alarm constraint $\beta$, the threshold $h$ can be set using $\beta=e^{-\omega_0 h}$ to ensure that $FAR \le \beta$. Hence, we get the equation $h=\frac{-\log \beta}{\omega_0}$, where we need to find $\omega_0$ from $E[e^{\omega_0 D_t}] = 1$. 

We firstly derive the asymptotic distribution of the anomaly evidence $D_t$ in the absence of anomalies. Its cumulative distribution function is given by 
\[
P(D_t \leq y) = P((L_t)^d \leq (L_{(M)})^d + y).
\]
It is sufficient to find the probability distribution of $(L_t)^d$, the $d$th power of the $k$NN distance at time $t$. Independent $d$-dimensional instances $\{\vx_t\}$ over time form a Poisson point process. The nearest neighbor ($k=1$) distribution for a Poisson point process is given by
\[
P(L_t \leq r) = 1 - \exp(-\Lambda(b(\vx_t,r)))
\]
where $\Lambda(b(\vx_t,r))$ is the arrival intensity (i.e., Poisson rate measure) in the $d$-dimensional hypersphere $b(\vx_t,r)$ centered at $\vx_t$ with radius $r$ \cite{chiu2013stochastic}. Asymptotically, for a large number of training instances as $N_2\to\infty$, under the null (nominal) hypothesis, the nearest neighbor distance $L_t$ of $\vx_t$ takes small values, defining an infinitesimal hyperball with homogeneous intensity $\lambda=1$ around $\vx_t$. Since for a homogeneous Poisson process the intensity is written as $\Lambda(b(\vx_t,r)) = \lambda |b(\vx_t,r)|$ \cite{chiu2013stochastic}, where $|b(\vx_t,r)| = \frac{\pi^{d/2}}{\Gamma(d/2+1)}r^d = v_d r^d$ is the Lebesgue measure (i.e., $d$-dimensional volume) of the hyperball $b(\vx_t,r)$, we rewrite the nearest neighbor distribution as
\begin{align*}
P(L_t \le r) &= 1-\exp\left( -v_d r^d \right),
\end{align*}
where $v_d = \frac{\pi^{d/2}}{\Gamma(d/2+1)}$ is the constant for the $d$-dimensional Lebesgue measure. 

Now, applying a change of variables we can write the probability density of $(L_t)^d$ and $D_t$ as
\begin{align}
f_{(L_t)^d}(y) &= \frac{\partial}{\partial y} \left[1-\exp\left( -v_d y \right) \right], \nn\\
&= v_d \exp(-v_d y), \nn\\
\label{eq:pdf}
f_{D_t}(y) &= v_d \exp(-v_d (L_t)^d) \exp(-v_d y)
\end{align}

Using the probability density derived in \eqref{eq:pdf}, $E[e^{\omega_0 D_t}] = 1$ can be written as
\begin{align}
1 &= \int_{-(L_{(M)})^d}^{\phi} e^{\omega_0 y} v_d e^{-v_d (L_t)^d} e^{-v_d y} dy, \nn\\
\frac{e^{v_d (L_{(M)})^d}}{v_d} &= \int_{-(L_{(M)})^d}^{\phi} e^{(\omega_0-v_d)y}dy, \nn\\
&= \frac{e^{(\omega_0-v_d)y}}{\omega_0-v_d}\Biggr|_{-(L_{(M)})^d}^{\phi}, \nn\\ 
&=\frac{e^{(\omega_0-v_d)\phi} - e^{(\omega_0-v_d)(-(L_{(M)})^d)}}{\omega_0-v_d},
\end{align}
where $-(L_{(M)})^d$ and $\phi$ are the lower and upper bounds for$D_t=(L_t)^d-(L_{(M)})^d$. The upper bound $\phi$ is obtained from the training set. 

As $N_2\to\infty$, since the $d$th power of $(1-\alpha)$th percentile of nearest neighbor distances in training set goes to zero, i.e., $(L_{(M)})^d \to 0$, we have 
\begin{align}
e^{(\omega_0-v_d)\phi} &= \frac{e^{v_d (L_{(M)})^d}}{v_d}(\omega_0-v_d) + 1. \nn
\end{align}

We next rearrange the terms to obtain the form of $e^{\phi x} = a_0(x+\theta)$ where $x=\omega_0-v_d$, $a_0=\frac{e^{v_d (L_{(M)})^d}}{v_d}$, and $\theta=\frac{v_d}{e^{v_d (L_{(M)})^d}}$. The solution for $x$ is given by the Lambert-W function \cite{scott2014asymptotic} as $x = -\theta - \frac{1}{\phi} \mathcal{W}(-\phi e^{-\phi\theta}/a_0)$, hence
\begin{align}
\omega_0 = v_d - \theta -\frac{1}{\phi} \mathcal{W}\left( -\phi \theta e^{-\phi\theta } \right). \nn
\end{align}

\bibliographystyle{IEEEtran}
\bibliography{References}